\documentclass[11pt,twoside,a4paper,cmspaper,final,collab]{cms-tdr}
\def\svnVersion{88bdc8b}\def\svnDate{2026/10/14}\def\cmsCernNoTag{CERN-EP-2026-099}\def\cmsCernDate{\today}\def\cmsMessage{Published in Physics Letters B as \href{http://dx.doi.org/10.1016/j.physletb.2026.140988}{\doi{10.1016/j.physletb.2026.140988}.}}
\begin{document}\cmsNoteHeader{SMP-23-007}

\providecommand{\cmsTable}[1]{\resizebox{\textwidth}{!}{#1}}
\ifthenelse{\boolean{cms@external}}{\providecommand{\cmsLeft}{upper\xspace}}{\providecommand{\cmsLeft}{left\xspace}}
\ifthenelse{\boolean{cms@external}}{\providecommand{\cmsRight}{lower\xspace}}{\providecommand{\cmsRight}{right\xspace}}

\providecommand{\PGgst}{\HepParticle{\PGg}{}{\star}\xspace}
\newcommand{\PGZ}{\ensuremath{\PGgst/\PZ}\xspace}
\newcommand{\ptll}{\ensuremath{\pt^{\Pell\Pell}}\xspace}
\newcommand{\yll}{\ensuremath{y^{\Pell\Pell}}\xspace}
\newcommand{\mll}{\ensuremath{m_{\Pell\Pell}}\xspace}
\newcommand{\ptmm}{\ensuremath{\pt^{\PGm\PGm}}\xspace}
\newcommand{\ymm}{\ensuremath{y^{\PGm\PGm}}\xspace}
\newcommand{\mmm}{\ensuremath{m_{\PGm\PGm}}\xspace}
\newcommand{\ptm}{\ensuremath{\pt^{\PGm}}\xspace}

\cmsNoteHeader{SMP-23-007}

\title{Measurement of the \texorpdfstring{$\PZ \to \PGmp\PGmm$}{Z to mu+mu-} angular coefficients in \texorpdfstring{$\Pp\Pp$}{pp} collisions at \texorpdfstring{$\sqrt{s} = 13\TeV$}{sqrt(s) = 13 TeV} as functions of transverse momentum and rapidity}

\author[cern]{the CMS Collaboration}

\date{\today}

\abstract{A measurement of the eight angular polarization coefficients, $A_0$ to $A_7$, in the cross section for the Drell--Yan production of two muons is presented. The analysis is based on proton-proton ($\Pp\Pp$) collision data recorded with the CMS detector at the LHC at a center-of-mass energy of $\sqrt{s}=13\TeV$, corresponding to an integrated luminosity of 140\fbinv. The coefficients are determined double differentially in eight intervals of transverse momentum and two intervals of rapidity of the muon pair $\PGmp\PGmm$. The results are presented for the $\PGmp\PGmm$ invariant mass range 81--101\GeV and are compared with theoretical predictions calculated at next-to-next-to-leading order in perturbative quantum chromodynamics. The measurement provides relevant information about the underlying partonic dynamics and the \PZ boson production mechanisms.}

\hypersetup{%
pdfauthor={CMS Collaboration},%
pdftitle={Measurement of the Z to mu+mu- angular coefficients in pp collisions at sqrt(s) = 13 TeV as functions of transverse momentum and rapidity},%
pdfsubject={CMS},%
pdfkeywords={CMS, electroweak, Z polarization, angular coefficients}}

\maketitle 

\section{Introduction}

Processes with dilepton final states play a key role in collider experiments, providing clean experimental signatures and precise probes of the underlying dynamics of particle interactions. Particularly, the Drell--Yan (DY) process ($\Pp\Pp \to \PGZ \to \Pellp\Pellm$, where $\Pp\Pp$ stands for proton-proton collisions and $\Pell$ represents the $\Pe$, $\PGm$, and $\PGt$ leptons)~\cite{Drell-Yan_1970} is sensitive to the electroweak (EW) \PGZ couplings and also provides unique information about lepton-hadron correlations~\cite{Mirkes_WandZPolarizEffects_1994, Mirkes_AngDecayDistrW-bosNLO_1992}. Such correlations are essential for understanding the dynamics of the initial-state partons and the effects of quantum chromodynamics (QCD), offering a window into parton distribution functions (PDFs)~\cite{Bodek_UsingAfbForPDF_2015, Zhang_EtractBoerMulder_2008, TMDandAC_2024} and their connection with the angular structure of the final states. Within the parton model, this connection is encoded in eight angular polarization coefficients $A_i$ ($i=0,\ldots,7$). 

We consider the DY process in the Collins--Soper (CS) frame~\cite{CollinsSoper_1977}, where the $z$ axis bisects the angle between the momentum vectors of the colliding protons, with the positive direction pointing closer to the momentum direction of the lepton pair, \ie of the \PGZ boson in the laboratory frame. The $x$ axis lies in the same plane as the momenta of the colliding protons, but orthogonal to the $z$ axis. The $y$ axis is then defined to complete the right-handed system. To be consistent with previous measurements~\cite{CMS_AiMeasur_8TeV_2015, ATLAS_AiMeasur_8TeV_2016}, the momentum of the beam proton closest in rapidity to the \PGZ boson is considered the ``target momentum" in our measurement.

\begin{figure}[ht]
\centering
\includegraphics[width=0.48\textwidth]{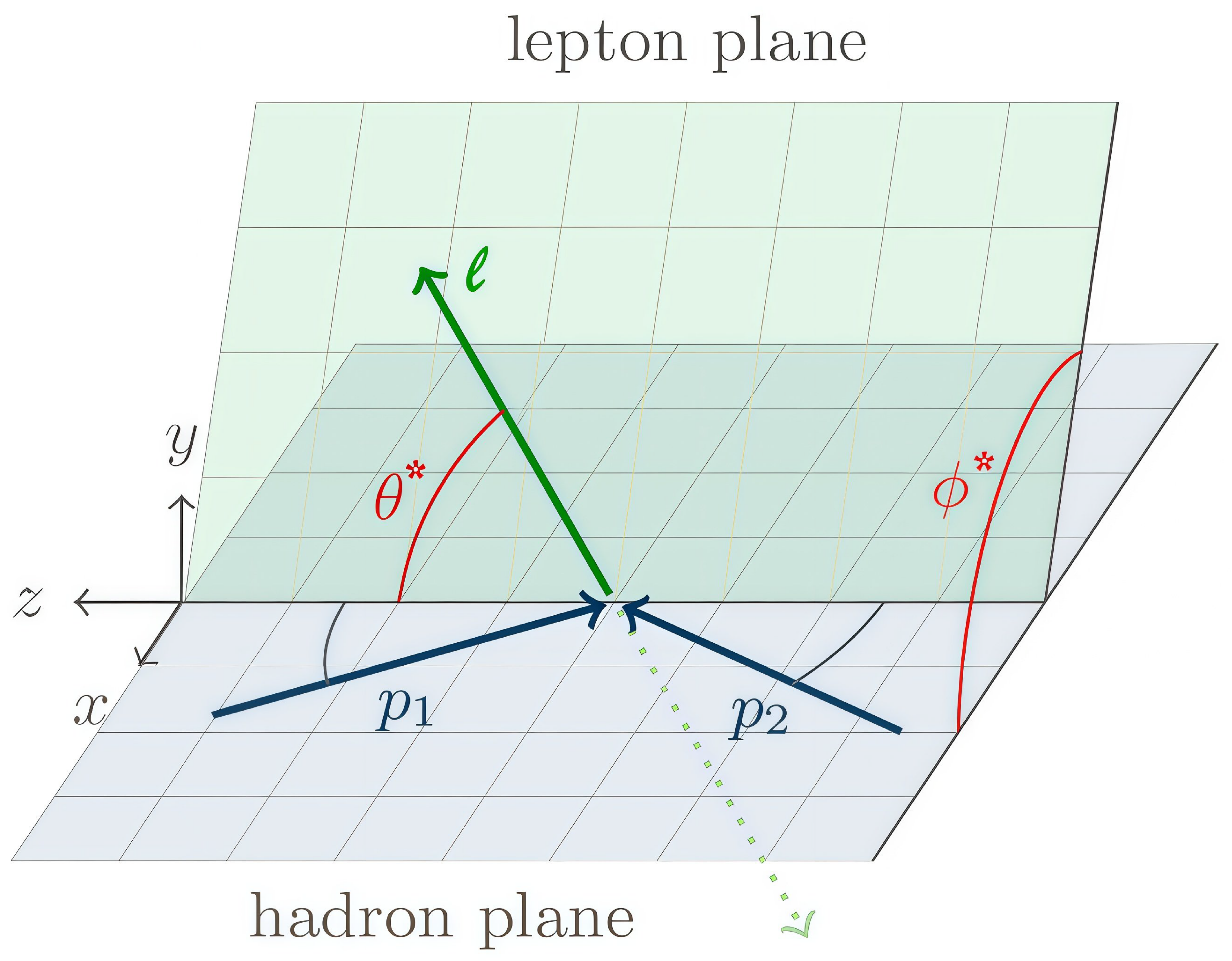}
\caption{The definition of the CS frame and angles $\phi^*$ and $\theta^*$ of the negatively charged lepton produced in the \PGZ decay. The $p_1$ and $p_2$ vectors indicate the directions of the incoming proton's momenta in the dilepton rest frame and $\Pell$ indicates the momentum of the negatively charged lepton~\cite{FigOfCollinsSoper_2017}.}
\label{fig_CollinsSoperFrame}
\end{figure}

The differential cross section depends on the azimuthal angle $\phi^*$ and polar angle $\theta^*$ as
\ifthenelse{\boolean{cms@external}}
{ 
\begin{equation}\begin{aligned}
&\frac{\rd^5\sigma}{\rd\mll\rd\yll\rd\ptll\rd\theta^*\rd\phi^*}
\propto
\frac{\rd^3\sigma^{\text{unp}}}{\rd\mll\rd\yll\rd\ptll}
\Bigl[
(1+\cos^2\theta^*) \\
&+ \frac{A_0}{2}(1-3\cos^2\theta^*) 
+ A_1\sin(2\theta^*)\cos\phi^* \\
&+ \frac{A_2}{2}\sin^2\theta^*\cos(2\phi^*)  
+ A_3\sin\theta^*\cos\phi^*   \\
&+ A_4\cos\theta^* 
+ A_5\sin^2\theta^*\sin(2\phi^*)\\
&+ A_6\sin(2\theta^*)\sin\phi^* 
+ A_7\sin\theta^*\sin\phi^*
\Bigr],
\end{aligned}
\label{DoubleDiffCrossSection}\end{equation}
}
{ 
\begin{equation}\begin{aligned}
&\frac{\rd^5\sigma}{\rd\mll\rd\yll\rd\ptll\rd\theta^*\rd\phi^*}
\propto\;
\frac{\rd^3\sigma^{\text{unp}}}{\rd\mll\rd\yll\rd\ptll}
\Bigl[
(1+\cos^2\theta^*)
+ \frac{A_0}{2}(1-3\cos^2\theta^*) \\
& + A_1\sin(2\theta^*)\cos\phi^* 
+ \frac{A_2}{2}\sin^2\theta^*\cos(2\phi^*)
+ A_3\sin\theta^*\cos\phi^*  \\
&+ A_4\cos\theta^* 
+ A_5\sin^2\theta^*\sin(2\phi^*)
+ A_6\sin(2\theta^*)\sin\phi^* 
+ A_7\sin\theta^*\sin\phi^*
\Bigr],
\end{aligned}
\label{DoubleDiffCrossSection}
\end{equation}
}
where $\sigma^{\text{unp}}$ is unpolarized cross section and Figure~\ref{fig_CollinsSoperFrame} illustrates the definition of the $\phi^*$ and $\theta^*$ angles in the CS frame.  

The angular polarization coefficients $A_i$ are functions of the lepton pair kinematic variables: transverse momentum \ptll, rapidity \yll, and invariant mass \mll. Importantly, these coefficients correspond to orthogonal angular terms in Eq.~(\ref{DoubleDiffCrossSection}), which define modulations in the angular distribution of the leptons. The coefficients encode contributions of different polarization states of the \PGZ boson reflecting processes occurring at the parton level. In calculations at leading order (LO) in QCD, all are expected to be close to zero at $\ptll \approx 0$, except for $A_4$, which describes the magnitude of asymmetry of the $\cos \theta^*$ distribution around zero. This phenomenon is also known as the forward-backward asymmetry $(A_{\text{FB}})$ and is a consequence of the parity violating nature of the EW decay of the \PGZ boson to a lepton pair. The coefficients $A_0$ and $A_2$ parametrize the longitudinal (L) polarization fraction of the \PGZ boson and the interference between its transverse (T) polarization states, respectively. The Lam--Tung relation $A_0 - A_2 = 0$~\cite{LumTung_1978} holds at leading order in perturbative QCD as a consequence of helicity conservation and the vector nature of the mediating boson. Deviations from this relation can occur because of higher-order QCD radiation, noncollinear parton dynamics, and other effects resulting from the intrinsic transverse momentum of the interacting partons~\cite{Faccioli_RotationInvariant_2010, Faccioli_ParticlePolarization_Book_2022, PENG_Teryaev_2016, Boer_IntrinsicTranverseMomentum1999}. The coefficient $A_1$ encodes the interference between longitudinal and transverse polarization states. The $A_3$ and $A_4$ coefficients are based on the axial and vector EW couplings of quarks and leptons. The coefficient $A_4$ is used for precise measurements of the squared sine of the effective EW mixing angle $\sin^2\theta_W^{\text{eff}}$~\cite{CDF_InderectSinThetaMeasur_1.96TeV_2013, CDF_FirstAiMeasur_1.96TeV_2011, WeakMixAngle_Meas_CMS_2025}. The $A_5$, $A_6$, and $A_7$ coefficients define the contribution of T-odd asymmetries to the total cross section and may be non-zero in one-loop processes at next-to-next-to-leading order (NNLO) in QCD~\cite{HAGIWARA_ToddAsym_1983, HAGIWARA_ProbOneLoop_1992}.

Some of the coefficients ($A_0$, $A_2$, $A_3$, and $A_4$) were measured for the first time in $\Pp\PAp$ collisions at a center-of-mass energy of 1.96\TeV at the Fermilab Tevatron by the CDF Collaboration~\cite{CDF_FirstAiMeasur_1.96TeV_2011}. More recent $A_i$ measurements were performed by the CMS~\cite{CMS_AiMeasur_8TeV_2015} and ATLAS~\cite{ATLAS_AiMeasur_8TeV_2016} Collaborations based on the LHC Run 1 data. The amount of data collected at $\sqrt{s} = 8$\TeV allowed measuring $A_0,A_1,A_2,A_3$, and $A_4$ in the \PZ pole region as functions of \PZ boson rapidity and transverse momentum, and made it possible to unequivocally establish a violation of the Lam--Tung relation. Moreover, ATLAS performed measurements using the combined $\EE$ and $\MM$ sample that indicated evidence of non-zero $A_5$, $A_6$, and $A_7$ with a significance of three standard deviations. In 2022, the LHCb Collaboration published the first measurement of the coefficients $A_0$ to $A_4$ in the dimuon channel in the forward rapidity region ($\ymm>2$) with dimuon transverse momentum $\ptmm<100$\GeV~\cite{LHCb_AiMeas_13TeV_2022}. To date, all the measurements agree with theoretical calculations at NNLO precision in QCD.

This Letter presents a measurement of the full set of angular polarization coefficients $A_0$ to $A_7$ in the DY dimuon channel process $(\Pp\Pp \to \PZ \to \MM)$ using LHC $\Pp\Pp$ collision data collected by the CMS experiment at $\sqrt{s} = 13$\TeV in 2016--2018. The data sample corresponds to an integrated luminosity of 140\fbinv. The results are compared with NNLO QCD predictions as a function of the dimuon rapidity and transverse momentum for $\abs{\ymm}<2.4$ and $\ptmm<400$\GeV.

\section{The CMS detector and event selection}
\label{sec_Selection}

The CMS apparatus~\cite{CMS:2008xjf} is a multipurpose, nearly hermetic detector, designed to trigger on~\cite{CMS:2020cmk,CMS:2016ngn} and identify electrons, muons, photons, and charged hadrons~\cite{CMS:2020uim,CMS:2018rym,CMS:2014pgm}. A global particle-flow (PF) algorithm~\cite{CMS:2017yfk} reconstructs all individual particle candidates in an event, combining information provided by the all-silicon tracker and by the crystal electromagnetic, and a brass and scintillator hadron calorimeters, operating inside a 3.8\unit{T} superconducting solenoid, with data from the gas-ionization muon detectors embedded in the flux-return yoke outside the solenoid. Muons are measured in the pseudorapidity range $\abs{\eta} < 2.4$, with detection planes made using three technologies: drift tubes, cathode strip chambers, and resistive plate chambers. The single-muon trigger efficiency exceeds 90\% over the full $\eta$ range, and the efficiency to reconstruct and identify muons is greater than 96\%. Matching muons to tracks measured in the silicon tracker results in a relative transverse momentum resolution, for muons with $\pt$ up to 100\GeV, of 1\% in the barrel and 3\% in the endcaps~\cite{CMS:2018rym}.

Events of interest are selected using a two-tiered trigger system. The first level (L1), composed of custom hardware processors, uses information from the calorimeters and muon detectors to select events at a rate of around 100\unit{kHz} within a fixed latency of 4\mus~\cite{CMS:2020cmk}. The second level, known as the high-level trigger (HLT), consists of a farm of processors running a version of the full event reconstruction software optimized for fast processing and reduces the event rate to a few kHz before data storage~\cite{CMS:2016ngn,CMS:2024aqx}.

For this analysis, events with two tracks reconstructed as \PGmp and \PGmm using the PF algorithm are selected. Both muons are required to be isolated from other particles originating from the same primary vertex, which means that the ratio of the scalar sum of $\pt$ of all other reconstructed physics objects within a cone $\Delta R = \sqrt{\smash[b]{(\Delta\phi)^2+(\Delta\eta)^2}}<0.4$ around the muon track to the muon $\ptm$ must be less than 0.15. Here, $\Delta\phi$ and $\Delta\eta$ are the distances in the $\phi$ and $\eta$ coordinates, respectively. The events are recorded using a single-muon requirement in the HLT, where at least one isolated muon track with $\ptm>24 $\GeV (for the 2016 and 2018 data taking periods) or $\ptm>27$\GeV (for the 2017 data taking period) and $\abs{\eta^{\PGm}}<2.4$ is reconstructed in the event. To reduce the $\PGgst$ contribution, the invariant mass of the dimuon must satisfy $81<\mmm<101$\GeV. The requirement $\abs{\ymm}<2.4$ is also applied because of the acceptance of the detector. Moreover, an additional requirement $\ptm > 26$ (20)\GeV for the leading (subleading) muon is applied for the 2016 and 2018 samples, and $\ptm > 29$ (20)\GeV for the 2017 sample. This selection is introduced to ensure an efficient and uniform response of the trigger in the \ptm region close to the trigger \pt threshold.

\section{Simulation and corrections}

To simulate the signal process, the \POWHEG v2.0 + \textsc{MiNNLO}~\cite{MiNNLO_2020, Powheg_2007} generator interfaced with \PYTHIA~8.240~\cite{Pythia8_2015} and \PHOTOS++~\cite{Photos_2015} is used.
 By interfacing with \PHOTOS++, quantum electrodynamics (QED) final state radiation (FSR) is considered at leading logarithmic precision, including matrix-element corrections and the effect of lepton pair production. The PDF set used is NNPDF3.1~\cite{NNPDF3.1_2017}  with the strong interaction coupling constant set to $\alpS(m_{\PZ}) = 0.118$, where $m_{\PZ}$ is the mass of the \PZ boson. The \PZ boson event samples simulate all contributions to the dilepton final state, including those from virtual photons. For the \PZ boson production, the ($G_{\PGm}$, $\sin^2\theta_W^{\text{eff}}$, $m_{\PZ}$) EW input scheme is used, where $G_{\PGm}$ is the Fermi constant extracted from the muon lifetime. Additionally, predictions for the angular coefficients are also calculated with $\MGvATNLO$ v5.2.2.2~\cite{Madgraph_2014} (with the NNPDF3.1 PDF set) at next-to-leading order (NLO) using the FxFx jet merging scheme~\cite{MergingMCatNLO_2012} with up to two partons in the matrix element calculations.

In this analysis, all background processes are modeled using NLO Monte Carlo (MC) event generators. The diboson processes ($\PW\PW$, $\PW\PZ$, and $\PZ\PZ$) are simulated with \POWHEG~\cite{Powheg_2007} and \MGvATNLO, while the $\PQt\PWpm$ process is generated with \POWHEG. The $\PW$+jets contribution is found to be negligible according to the \MGvATNLO prediction. The \MGvATNLO generator is also used to simulate the $\textrm{DY}\to\PGtp\PGtm$ and \ttbar processes for the 2017 and 2018 data sets. For the 2016 data set, the \ttbar process is simulated with \POWHEG~\cite{Powheg_2007} interfaced with \HERWIG~\cite{Herwig7_2020} using the CH3 tune~\cite{CMS_HERWIG7_CH3_2021}. For all other simulated samples in this work, parton showering, hadronization, the underlying event, and QED final-state radiation are modeled with \PYTHIA~8 using the CP5 tune~\cite{CMS_PYTHIA8_tunes_2020}, with the hard primordial-\kt parameter set to 2.225\GeV, as determined from a dedicated optimization using the \ptmm measurement of Ref.~\cite{ZbosXsec_2019}. Multiple $\Pp\Pp$ interactions in the same or neighboring bunch crossings (pileup) are simulated with \PYTHIA~8. Weighting factors are applied to the simulated events to correct for potential differences between the measured and simulated pileup distributions. Finally, all generated events are processed through a \GEANTfour-based detector simulation~\cite{GEANT4_2003}.

The background-to-signal ratio increases from approximately 0.03\% at $\ptmm<10$\GeV to about 2.9\% at $\ptmm>200$\GeV. The background from dimuon final states produced in diboson processes ($\PW\PW$, $\PW\PZ$, and $\PZ\PZ$) is dominant over the phase space of the measurement, contributing up to 2.8\% of the selected events at high \ptmm ($>200$\GeV). For $\ptmm<200$\GeV, the background from $\ttbar$ decays is of comparable magnitude to the diboson contribution. In the low \ptmm region ($<20$\GeV), the $\ttbar$ background becomes about 0.003--0.03\%, comparable to the contribution from muons originating from \PGt lepton decays in the $\text{DY} \to \PGtp\PGtm$ process.  

To account for the differences in the efficiencies between data and simulation, scale factors are derived and applied on an event-by-event basis. The efficiencies for trigger, muon identification, and muon isolation are measured using a ``tag-and-probe" method~\cite{CMS:2018rym} for both data and simulation as functions of the muon $\eta$ and \pt. The application of the efficiency correction can modify the event weight by up to 10\%. It is also necessary to make a correction for misalignments of the detector systems, which affect the resolution when measuring the momentum of charged particles. Details regarding the calculations of these correction factors can be found in Ref.~\cite{RochesterCorrection_2012}.

In a small fraction of the collected events, an imperfect synchronization of the L1 trigger (associated with the limited time resolution of the muon detectors) resulted in an incorrect assignment of the bunch crossing, so that some tracks were associated with the previous bunch crossing. The probability of this ``prefiring" effect does not exceed 1.5\%~\cite{CMSTrigPerform_2021} and depends on the muon $\eta$ and \pt. It is measured differentially in these variables to properly model this dependence. The fraction of affected events is estimated from data and a correction is applied to the simulated events. 

To avoid discrepancies with data in the shape of the \PZ boson kinematic distributions related to the model used by the MC generator, simulated events are reweighted in eight bins of \ptmm and two bins of \ymm (see binning in Section~\ref{sec_ExtractionOfAi}). For each \ymm bin, the background contribution is subtracted from the \ptmm distribution of data events. Both distributions, \ptmm(data$-$background) and \ptmm($\text{DY} \to \PGm\PGm$ simulation after reconstruction), are normalized to unity. Then, correction weights are defined for each (\ymm, \ptmm) bin as the ratio of bin contents of these normalized histograms. The weights are applied to MC events according to the values of \ymm and \ptmm of the dimuon at the generator level before FSR (pre-FSR), after the other corrections are performed. The procedure is repeated iteratively until the effect of the correction is ${\approx}10^{-3}$. The effect of the reweighting is most significant for \ptmm bins up to 35 \GeV and can reach 6\% of the bin content.

\begin{figure}[ht]
\centering
\ifthenelse{\boolean{cms@external}}
{ 
\includegraphics[width=0.49\textwidth]{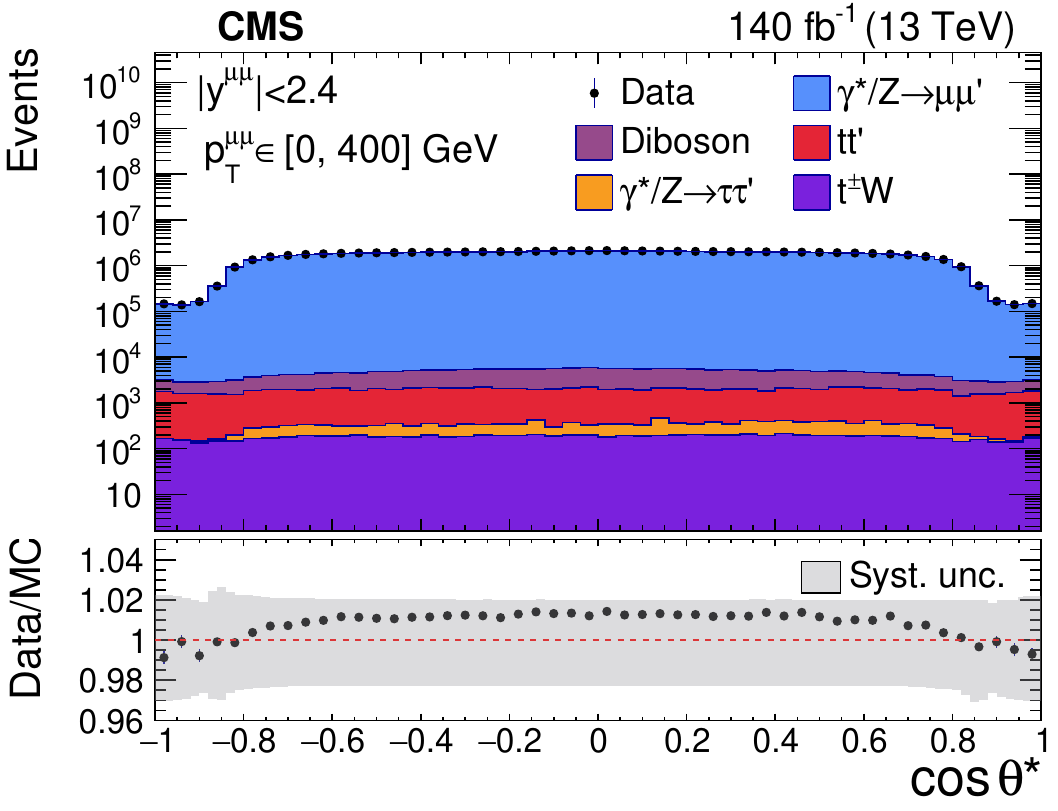}
\includegraphics[width=0.49\textwidth]{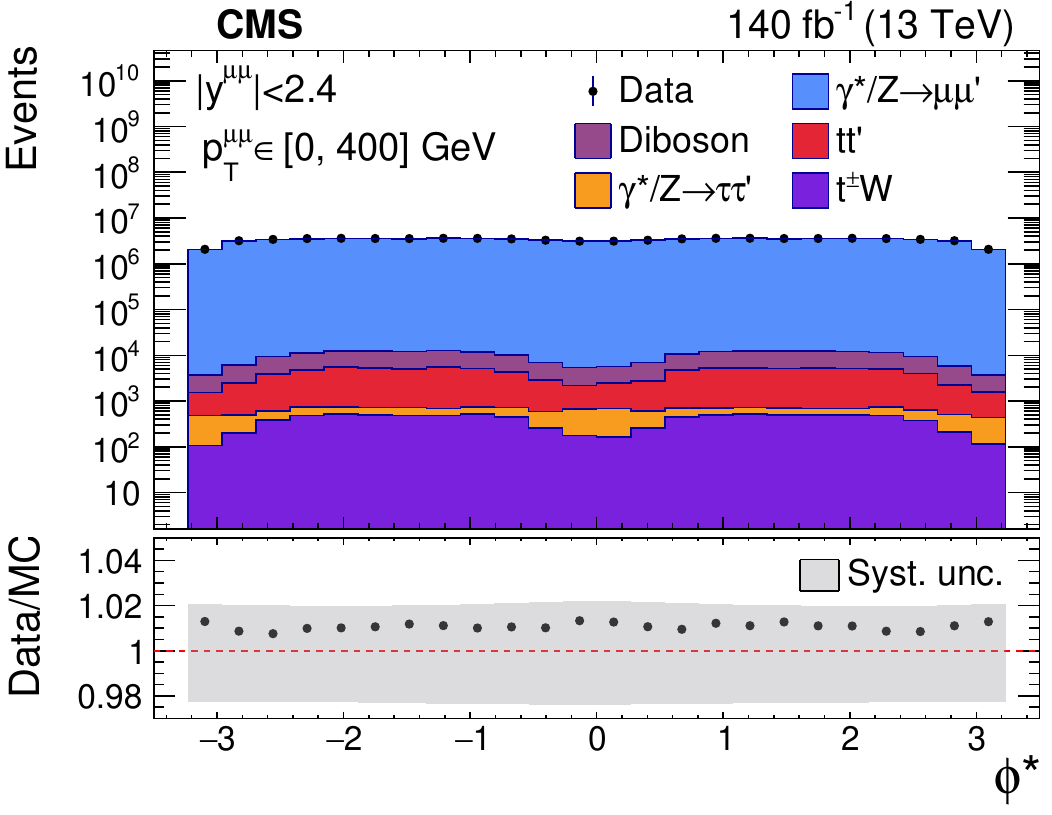}
}{
\includegraphics[width=0.45\textwidth]{Figure_002-a.pdf}
\includegraphics[width=0.45\textwidth]{Figure_002-b.pdf}
}
\caption{Distributions of events as functions of $\cos{\theta^*}$ (\cmsLeft) and $\phi^*$ (\cmsRight), averaged over \ptmm and \ymm. The measured distributions are represented by the black markers. The simulated contributions (from the signal and background processes) are shown by the colored histograms. The data/MC ratios are presented in the lower panels. The gray bands around unity represent the total systematic uncertainties.}
\label{figure_coscs_phics}
\end{figure}

Distributions as functions of $\cos{\theta^*}$ and $\phi^*$ for data and MC events, after all mentioned corrections are applied, are presented in Fig.~\ref{figure_coscs_phics}. Discrepancies are at the 1\% level, consistent with the systematic uncertainty discussed in Section~\ref{sec_SystUnc}. 

\section{Extraction of \texorpdfstring{$A_i$}{Ai}} 
\label{sec_ExtractionOfAi}         

The angular coefficients are measured in two bins of the dimuon absolute rapidity, $\abs{\ymm}=0$--1 and 1--2.4, and eight bins of \ptmm with edges at 0, 10, 20, 35, 55, 80, 120, 200, and 400\GeV.

At the pre-FSR level, the angular distribution of dimuon events in the CS frame $\mathrm{D}(\cos\theta^*,\phi^*)$ is described by Eq.~(\ref{DoubleDiffCrossSection}). However, the measured distribution is constructed from reconstructed final-state muons passing the event selection described in Sec.~\ref{sec_Selection}. Consequently, the observed angular distribution $\mathrm{D}(\cos\theta^*_{\text{obs}},\phi^*_{\text{obs}})$ is affected by final-state radiation, detector acceptance and resolution, and migration of events between neighboring \ptmm bins. Therefore, the observed distribution cannot be described directly by Eq.~(\ref{DoubleDiffCrossSection}), and a template method is employed to extract the angular coefficients.

In each $(\ptmm, \abs{\ymm})$ interval, Eq.~(\ref{DoubleDiffCrossSection}) can be written as
\begin{equation}
\frac{\rd^2\sigma}{\rd\theta^*\rd\phi^*}
\propto
\sum_{i=0}^{8}
P_i,f_i(\theta^*,\phi^*),
\end{equation}
where $f_i(\theta^*,\phi^*)$ are the trigonometric polynomials (basis function) corresponding to the angular coefficients in Eq.~(\ref{DoubleDiffCrossSection}). The parameters $P_i$ are related to the angular coefficients through $A_i=P_i/P_8$, where $P_8$ corresponds to the unpolarized cross section proportional to $(1+\cos^2\theta^*)$.
For each basis function $f_i(\theta^*,\phi^*)$, a detector-level template $T_i^{\mathrm{MC}}(\cos\theta^*_{\text{obs}},\phi^*_{\text{obs}})$ is constructed from simulated events. The intrinsic angular coefficients of the generator model, $P_i^{\text{ref}}$, are determined from angular moments following Ref.~\cite{Mirkes_WandZPolarizEffects_1994}. Each simulated event is assigned a weight
\begin{equation}
w_i =
\frac{
f_i(\theta^*,\phi^*)
}{
\sum_{j=0}^{8}
P_j^{\text{ref}}
f_j(\theta^*,\phi^*)
},
\label{eq_templateWeight}
\end{equation}
The denominator removes the dependence on the polarization model implemented in the event generator, while the numerator projects the event onto the corresponding basis function $f_i(\theta^*,\phi^*)$. The weights are computed from the pre-FSR kinematics, whereas the templates are filled using the corresponding reconstructed events after the full detector simulation, event reconstruction, and application of all analysis corrections. The resulting detector-level angular distributions define the templates $T_i^{\mathrm{MC}}(\cos\theta^*_{\text{obs}},\phi^*_{\text{obs}})$.

Under the assumption that final-state radiation and detector effects are accurately modeled by simulation, the measured angular distribution $\mathrm{D}^{\text{data}}(\cos\theta^*_{\text{obs}},\phi^*_{\text{obs}})$ can be represented as
\ifthenelse{\boolean{cms@external}}
{
\begin{multline}
\mathrm{D}^{\text{data}}
(\cos\theta^*_{\text{obs}},
\phi^*_{\text{obs}})
\approx \\
\sum_{i=0}^{8}
P_i\,
T_i^{\mathrm{MC}}
(\cos\theta^*_{\text{obs}},
\phi^*_{\text{obs}})
+
\mathrm{D}^{\text{bkg}}
(\cos\theta^*_{\text{obs}},
\phi^*_{\text{obs}}),
\label{eq_ForFit}
\end{multline}
}
{
\begin{equation}
\mathrm{D}^{\text{data}}
(\cos\theta^*_{\text{obs}},
\phi^*_{\text{obs}})
\approx
\sum_{i=0}^{8}
P_i\,
T_i^{\mathrm{MC}}
(\cos\theta^*_{\text{obs}},
\phi^*_{\text{obs}})
+
\mathrm{D}^{\text{bkg}}
(\cos\theta^*_{\text{obs}},
\phi^*_{\text{obs}}),
\label{eq_ForFit}
\end{equation} 
}
where $\mathrm{D}^{\text{bkg}}(\cos\theta^*_{\text{obs}},\phi^*_{\text{obs}})$ denotes the contribution from background processes.

The coefficients $P_i$ are extracted using a two-dimensional binned maximum-likelihood fit. Distributions $\mathrm{D}^{\text{data}}(\cos\theta^*_{\text{obs}},\phi^*_{\text{obs}})$, and $\mathrm{D}^{\text{bkg}}(\cos\theta^*_{\text{obs}},\phi^*_{\text{obs}})$, and all templates $T_i^{\mathrm{MC}}(\cos\theta^*_{\text{obs}},\phi^*_{\text{obs}})$ are divided into $14\times14=196$ bins of equal size in $(\cos\theta^*_{\text{obs}},\phi^*_{\text{obs}})$. The fit is performed with MINUIT~\cite{MINUIT_1994} using Poisson likelihoods independently in each $(\ptmm, \abs{\ymm})$ interval.

Because of the finite detector acceptance, the templates are not strictly orthogonal, leading to correlations among the fitted parameters. The largest correlations are observed between $P_0$, $P_2$, and $P_8$, reaching values of about 0.7 in some kinematic regions, while correlations involving the remaining coefficients are typically below 0.2.

The quality of the fit is assessed using the Pearson $\chi^2$ test, including both statistical and systematic uncertainties (Sec.~\ref{sec_SystUnc}). Across all kinematic intervals, the $\chi^2$ per degree of freedom is in the range 0.9--1.2, indicating a good description of the data.

The extraction procedure is validated using the $\mathrm{D}^{\mathrm{MC}}(\cos\theta_{\text{obs}}^*,\phi_{\text{obs}}^*)$ distribution of simulated events instead of $\mathrm{D}^{\text{data}}(\cos\theta^*_{\text{obs}},\phi^*_{\text{obs}})$ in Eq.~\ref{eq_ForFit}. The fitted coefficients are found to be consistent with the reference values within the statistical precision of the comparison, demonstrating that the effects of final-state radiation, detector acceptance, and event migration are well accounted by the template method. The fraction of events reconstructed in a \ptmm bin different from their true bin is below 7\% for the binning used in the measurement and has a negligible impact on the extracted coefficients.

\section{Systematic uncertainties}
\label{sec_SystUnc}

To estimate the systematic uncertainty in the measurement, each source of a systematic effect is independently varied within its own uncertainty and the corresponding variation in the $A_i$ value is evaluated. The difference between the varied $A_i$ value and its baseline value is taken as the uncertainty from the considered source. The following effects are taken into account: experimental uncertainties (pileup reweighting, efficiency of trigger, efficiencies of muon identification and isolation, detector misalignment, luminosity measurement uncertainty~\cite{Precisionluminositymeasurement_2021, CMS_Lumi_PAS_2025}, and prefiring), theoretical cross section uncertainties for signal and background processes, uncertainties from the limited number of events in the MC samples, and the shape of \ptmm distribution. The uncertainty related to the chosen PDF and the strong coupling $\alpS(m_{\PZ})$ value is evaluated following the PDF4LHC prescription~\cite{PDF4LHC_2016} using a set of symmetric Hessian eigenvectors including $\alpS(m_{\PZ})$ variations. To estimate the effect of missing higher orders in perturbative QCD, the factorization scale $\mu_\text{F}$ and the renormalization scale $\mu_\text{R}$ are varied independently by factors of 0.5 and 2, excluding the extreme case where both scales are varied in opposite directions. The maximum variation of the predicted coefficients from these scale variations is taken as the uncertainty. The uncertainties from the latter two sources are also calculated for the reported MC predictions and found to change the values of the predicted coefficients by less than $10^{-4}$. 

The contribution of an $A_i$ uncertainty source can vary depending on the coefficient and the kinematic range. The most significant uncertainties for $A_0, A_1, A_2$ come from the efficiency corrections. The size of the simulated event sample contributes significantly to the all coefficient uncertainties for $\ptmm > 80$\GeV, reaching the value of 97\% from total systematic uncertainty on the $A_6$ coefficient in the bin $\ptmm \in [120,200] \GeV$. The coefficients $A_3$ and $A_4$ are also sensitive to the PDF$+\alpS(m_{\PZ})$ uncertainty. The total systematic uncertainty is calculated as the sum in quadrature of the individual uncertainties, separately for positive and negative effects. The uncertainty in the $A_0-A_2$ difference, that tests the Lam--Tung relation, is calculated as the quadrature sum of the $A_0$ and $A_2$ uncertainties with the correlation between these coefficients taken into account.

\section{Results}

\begin{figure*}[htbp!]
\centering
\includegraphics[width=0.42\textwidth]{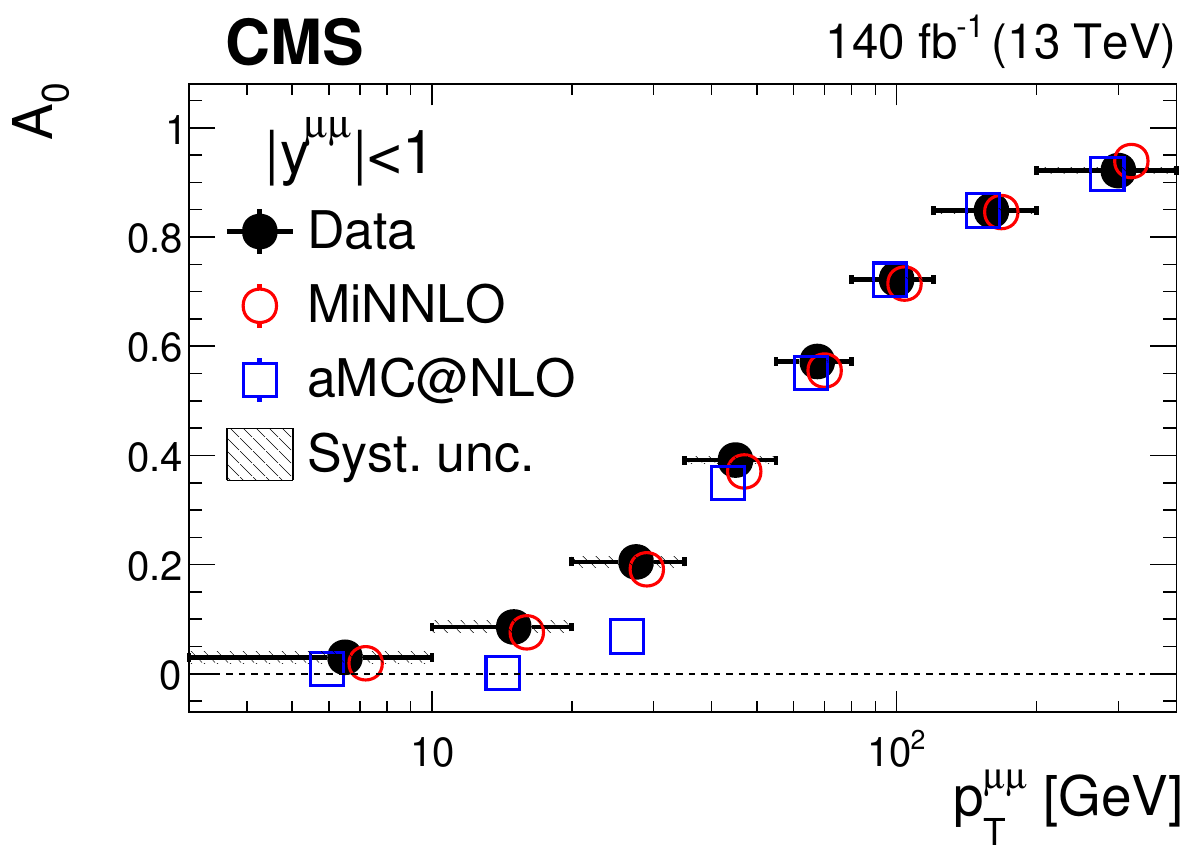}
\includegraphics[width=0.42\textwidth]{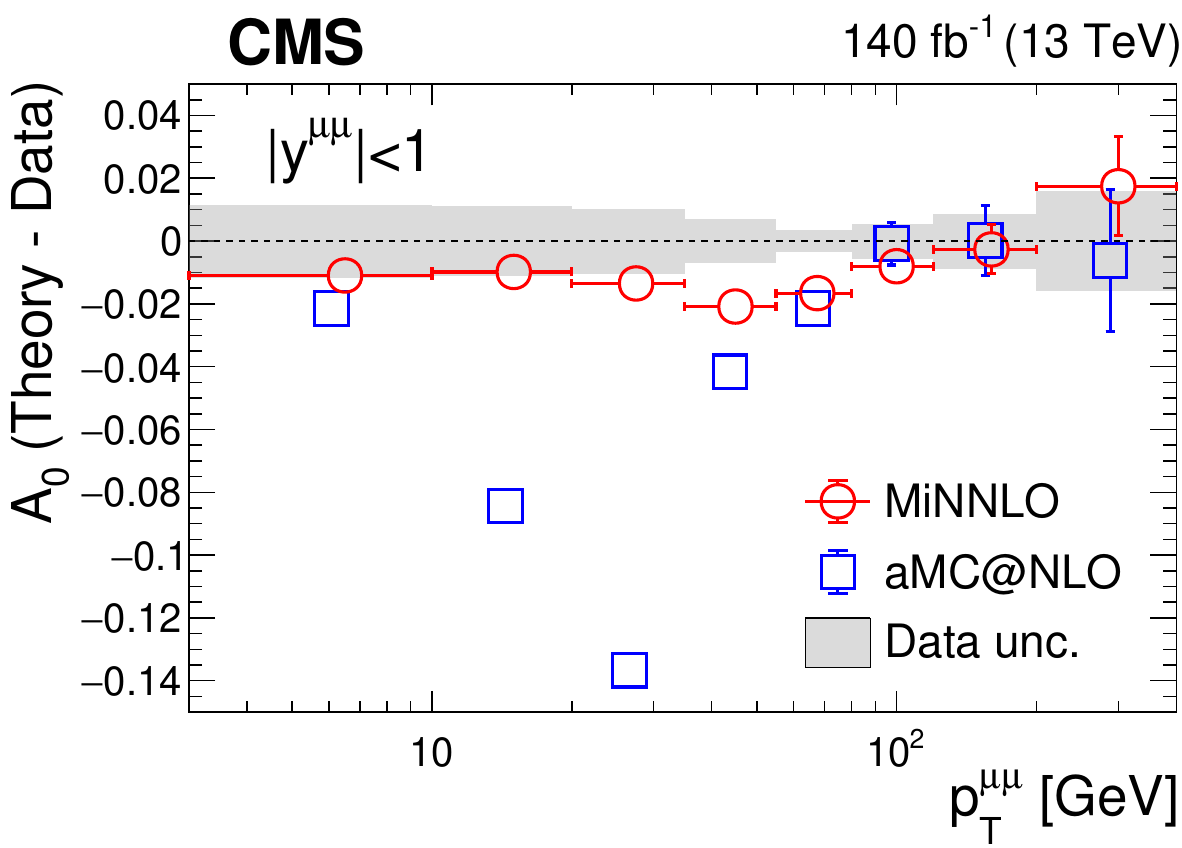}
\includegraphics[width=0.42\textwidth]{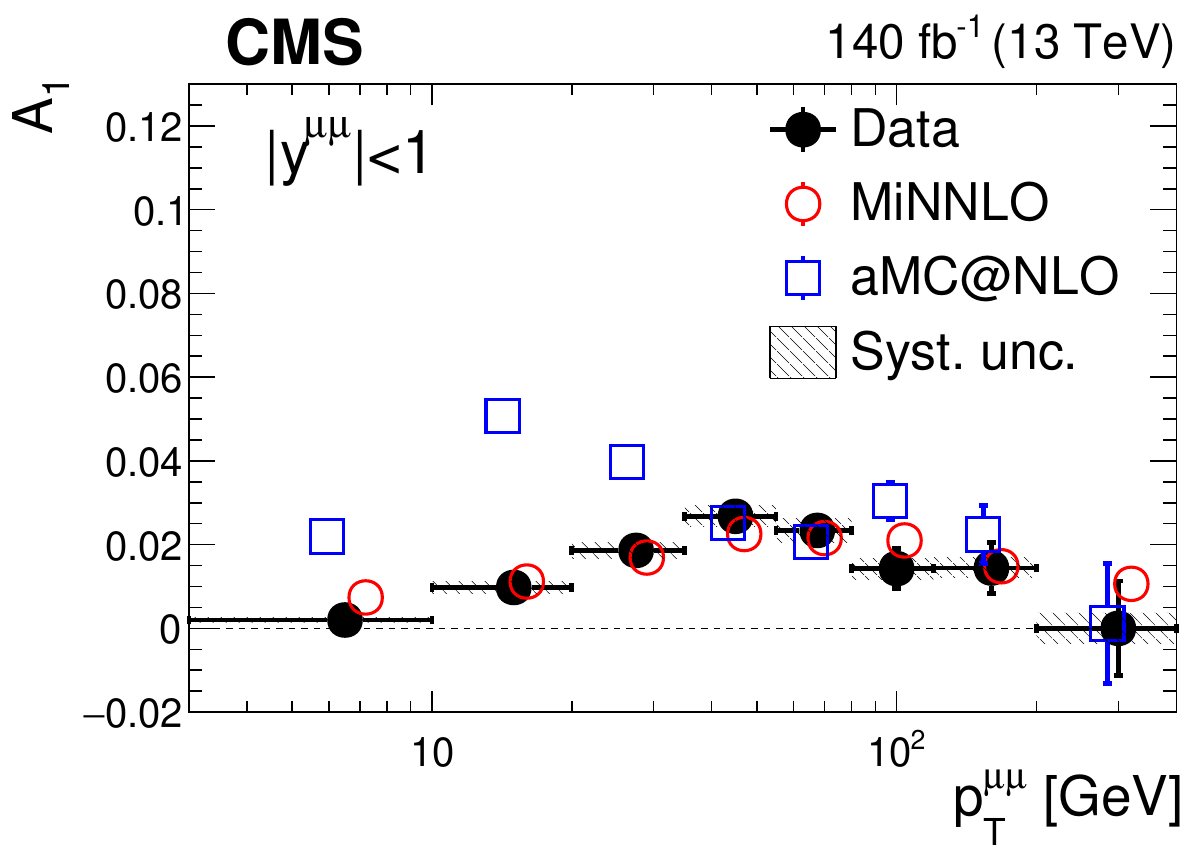}
\includegraphics[width=0.42\textwidth]{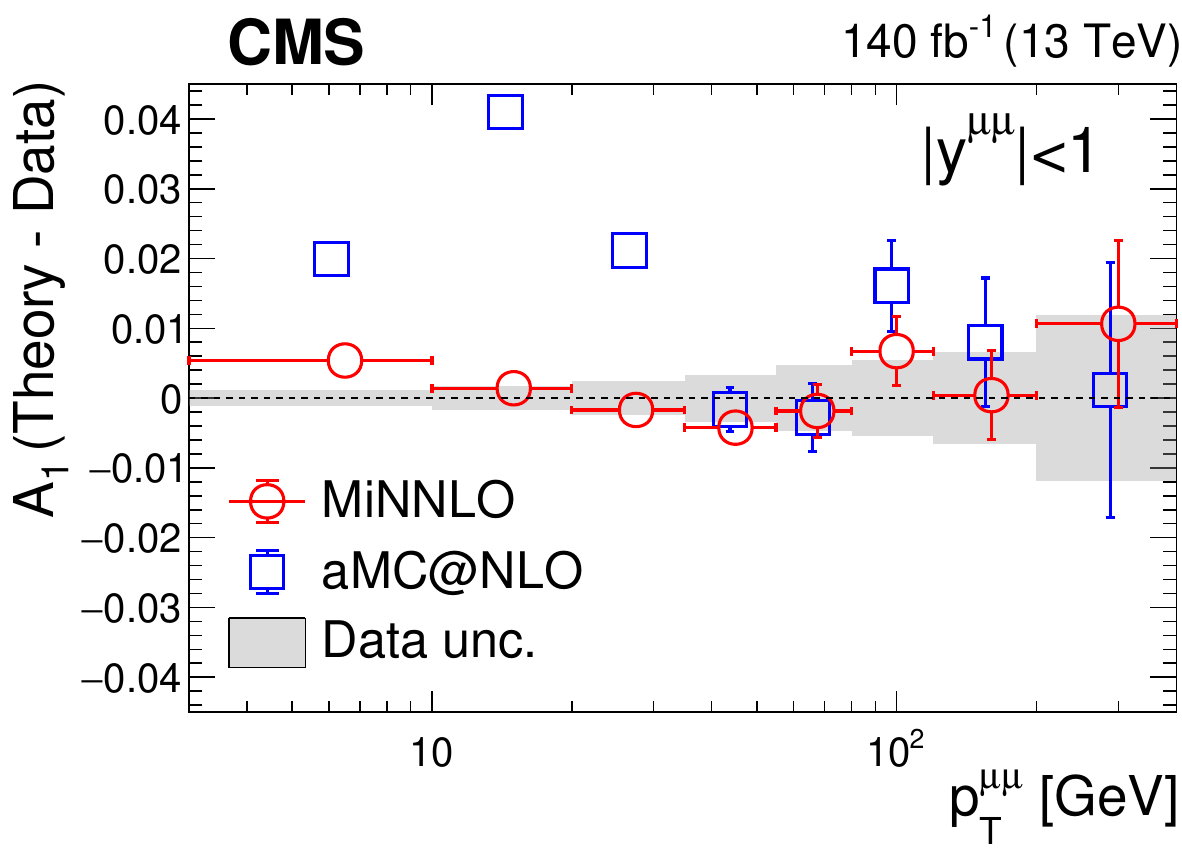}
\includegraphics[width=0.42\textwidth]{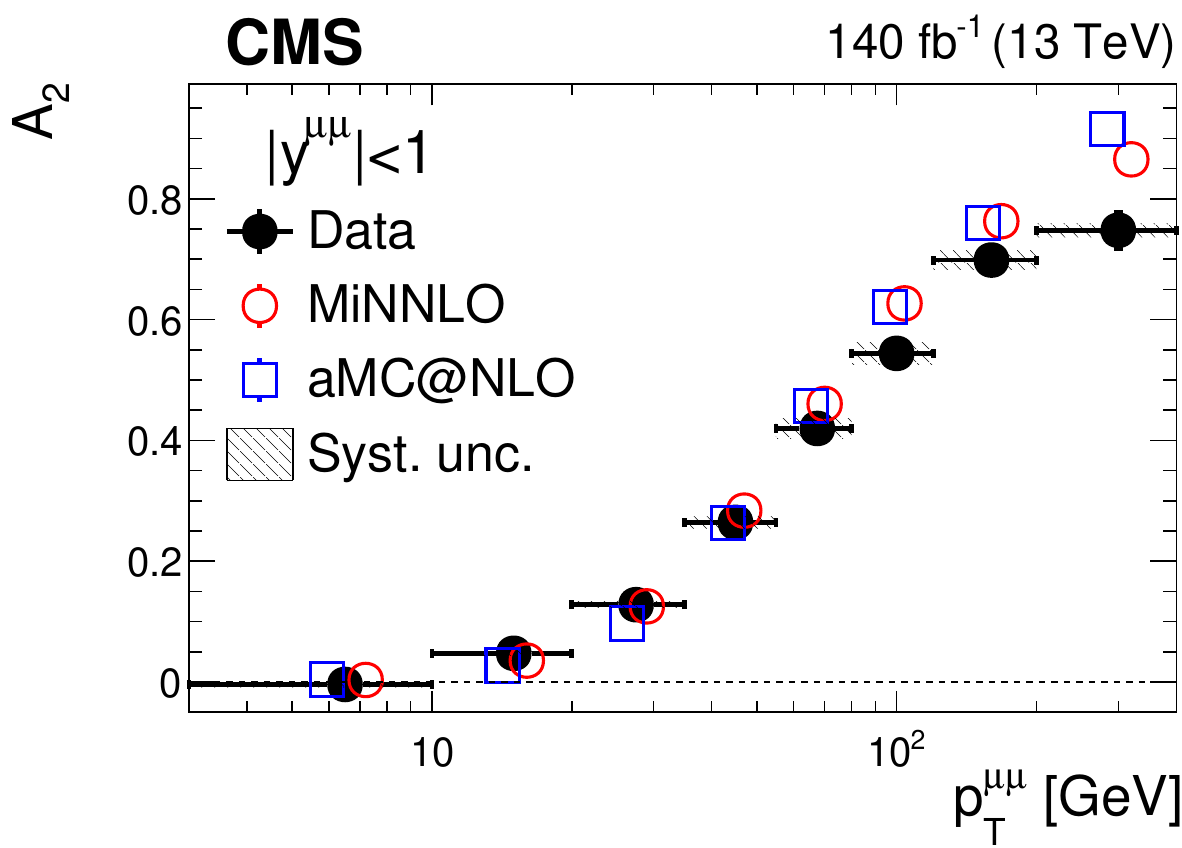}
\includegraphics[width=0.42\textwidth]{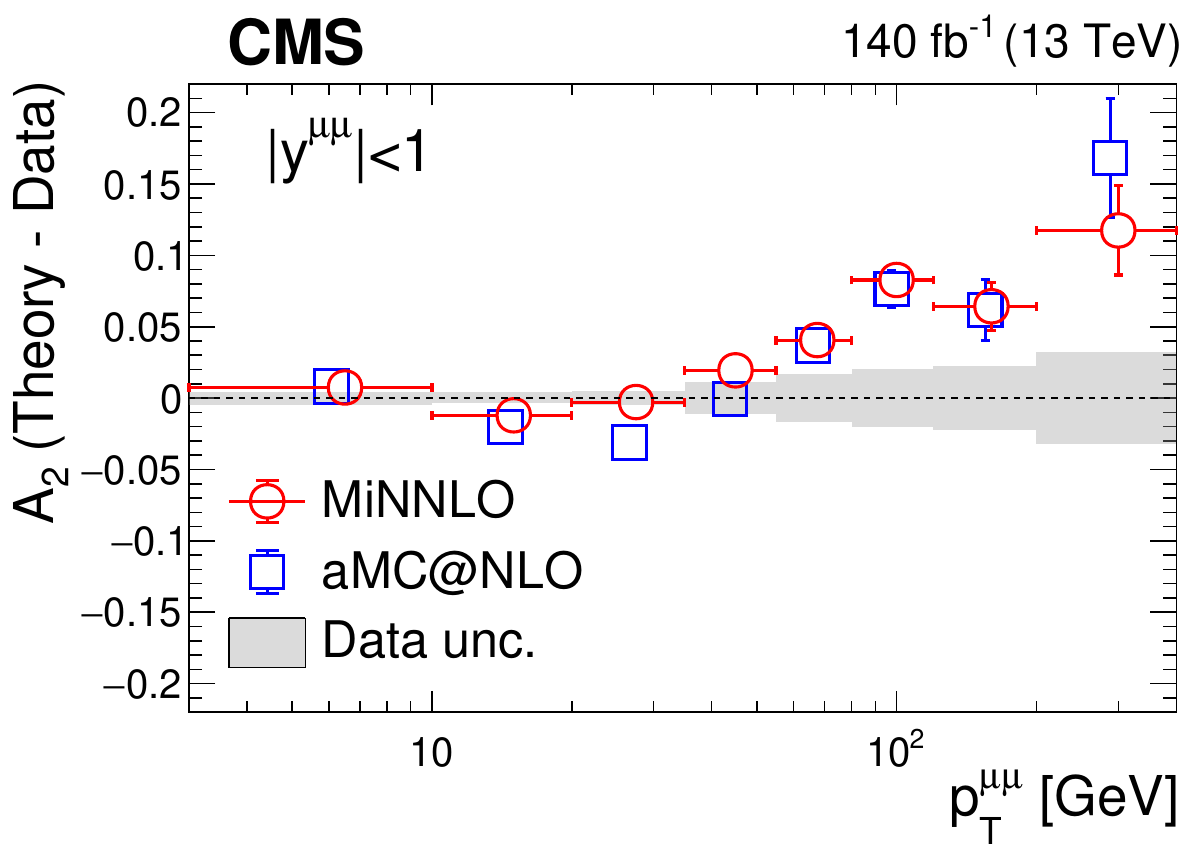}
\includegraphics[width=0.42\textwidth]{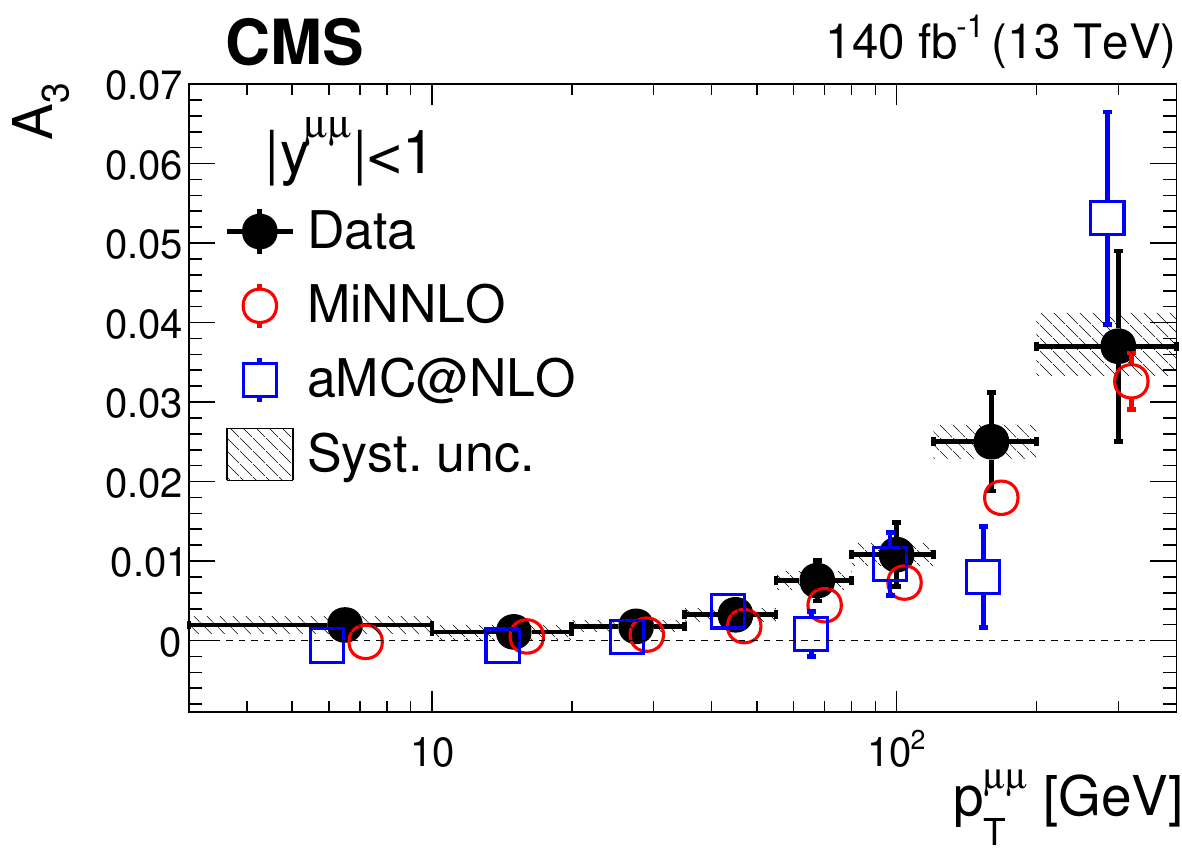}
\includegraphics[width=0.42\textwidth]{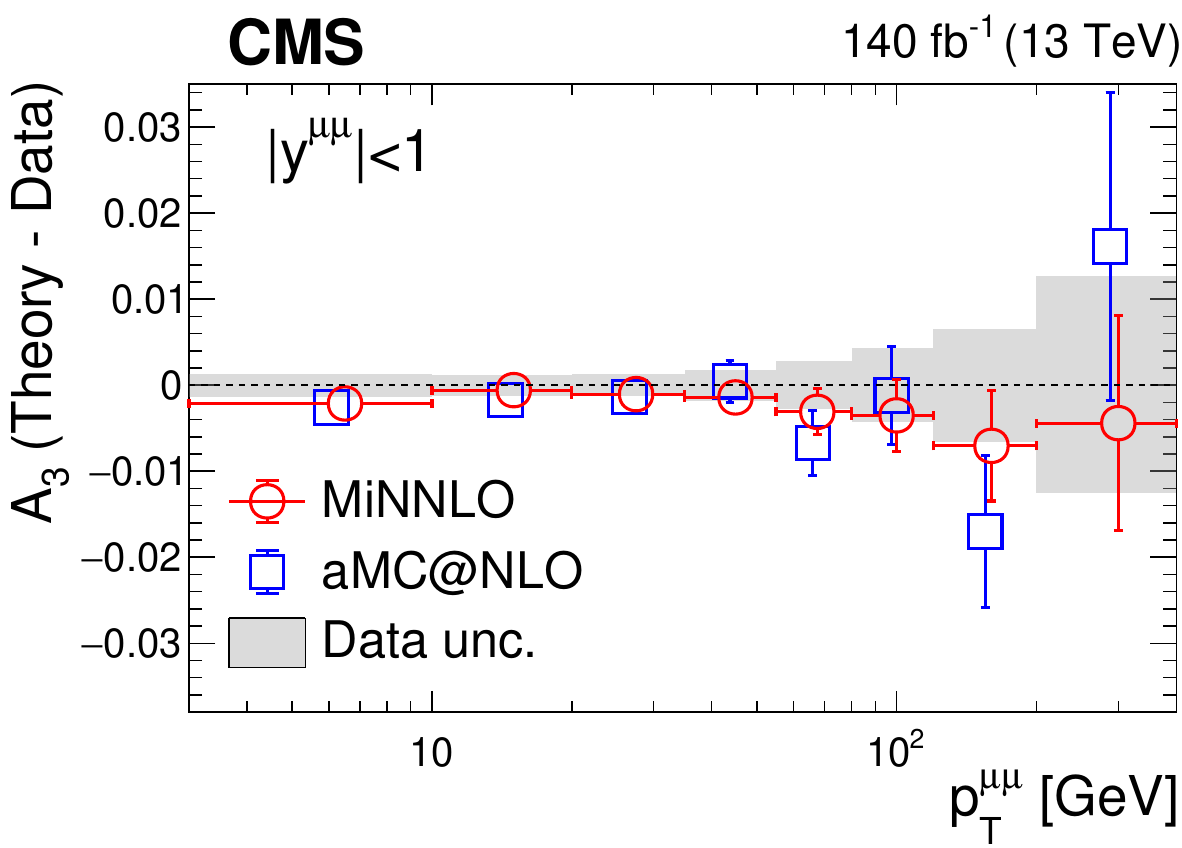}
\caption{Left: Polarization coefficients $A_{0}$ to $A_{3}$ measured in the CS frame in bins of \ptmm for $\abs{\ymm} < 1$. The data points are shown as black circles. The \POWHEG\!+\textsc{MiNNLO} and the \MGvATNLO predictions are represented by the red circles and blue squares, respectively, slightly displaced horizontally for improved visibility. The vertical bars (hatched boxes) represent the statistical (systematic) uncertainties. The theoretical uncertainties of the predictions are negligible and not shown. Right: Difference between the predicted and measured values. The gray area around zero represents the total uncertainty of the measurement, while the vertical bars represent the statistical uncertainties of the predictions.}
\label{CoeffsWithUncert_A0A3_Y0}
\end{figure*}

\begin{figure*}[htbp!]
\centering 
\includegraphics[width=0.42\textwidth]{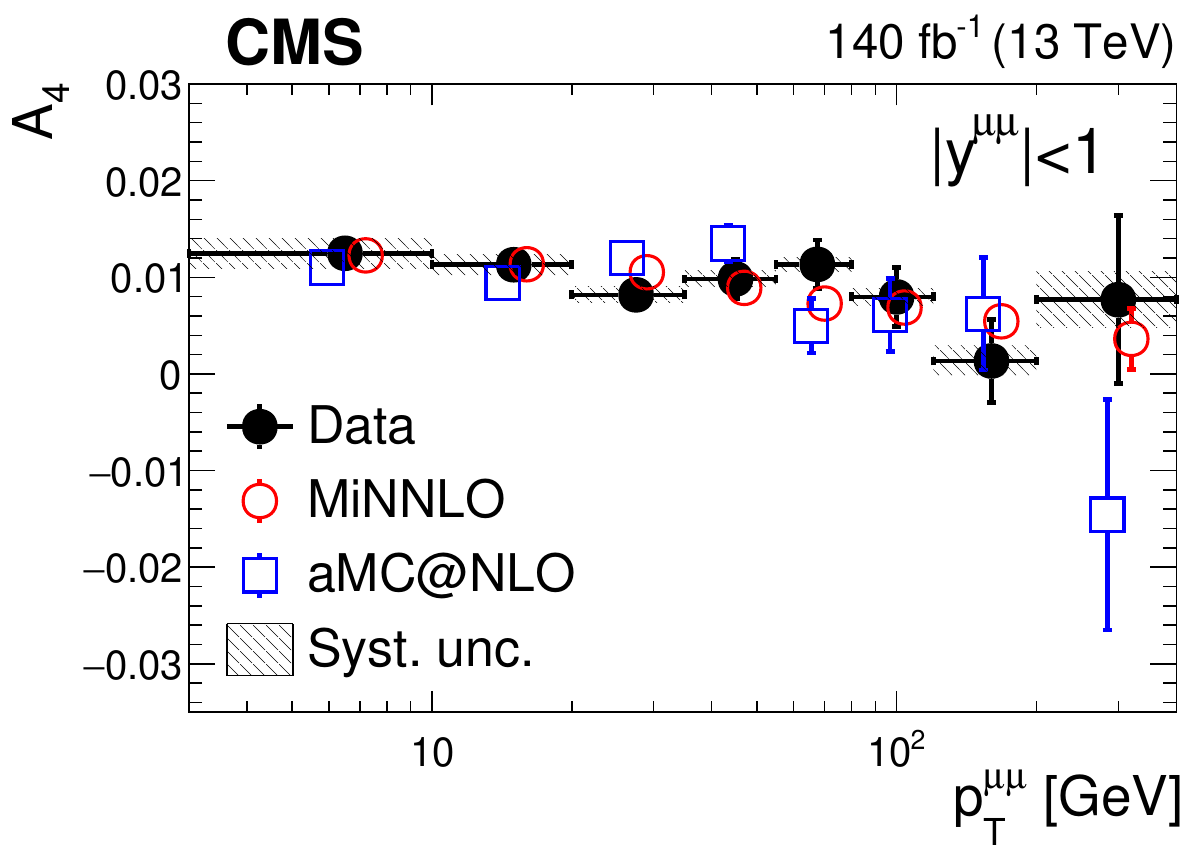}
\includegraphics[width=0.42\textwidth]{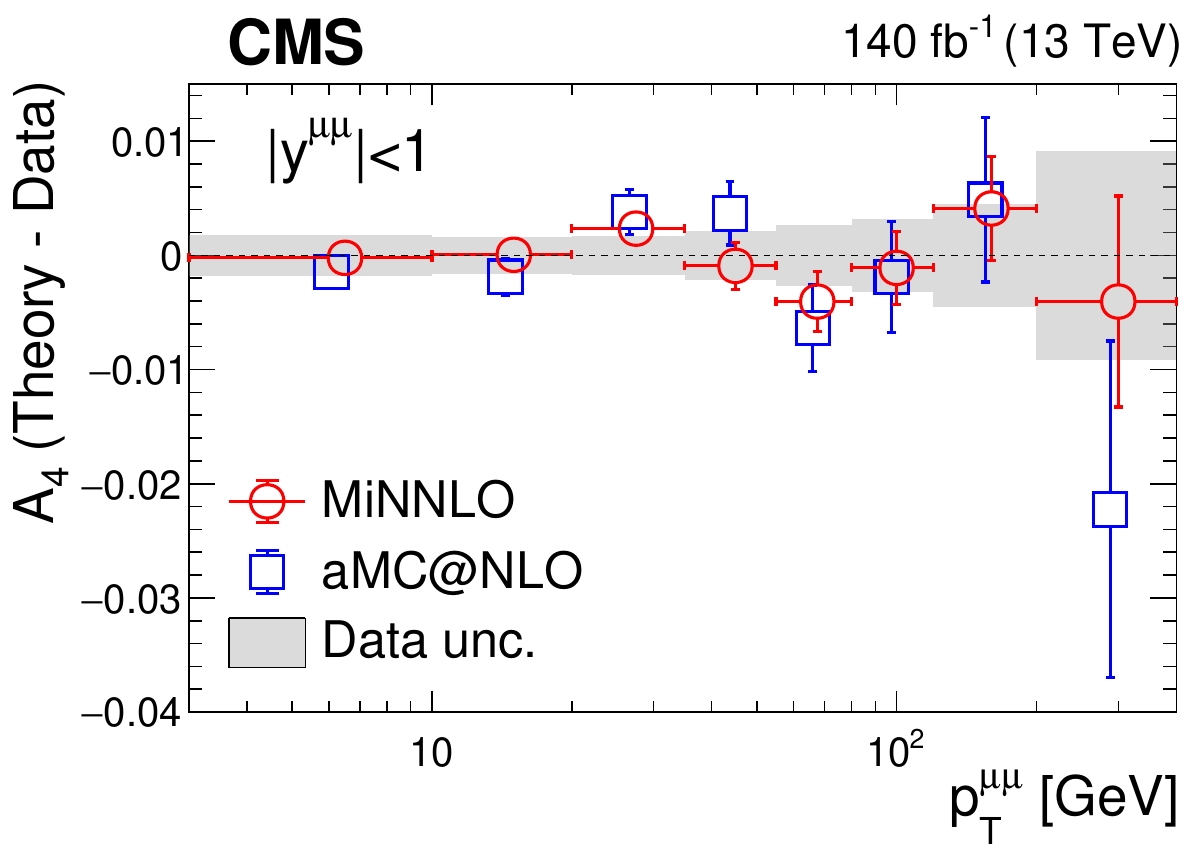}
\includegraphics[width=0.42\textwidth]{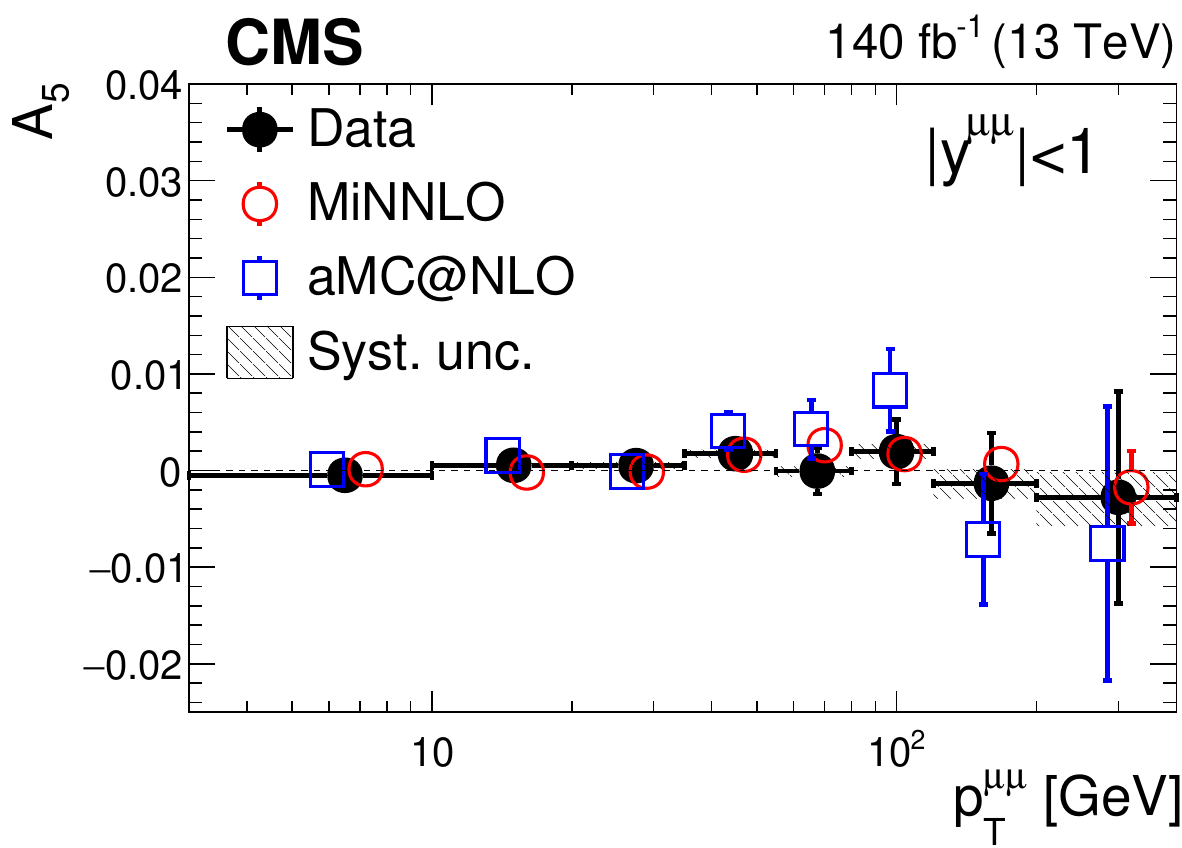}
\includegraphics[width=0.42\textwidth]{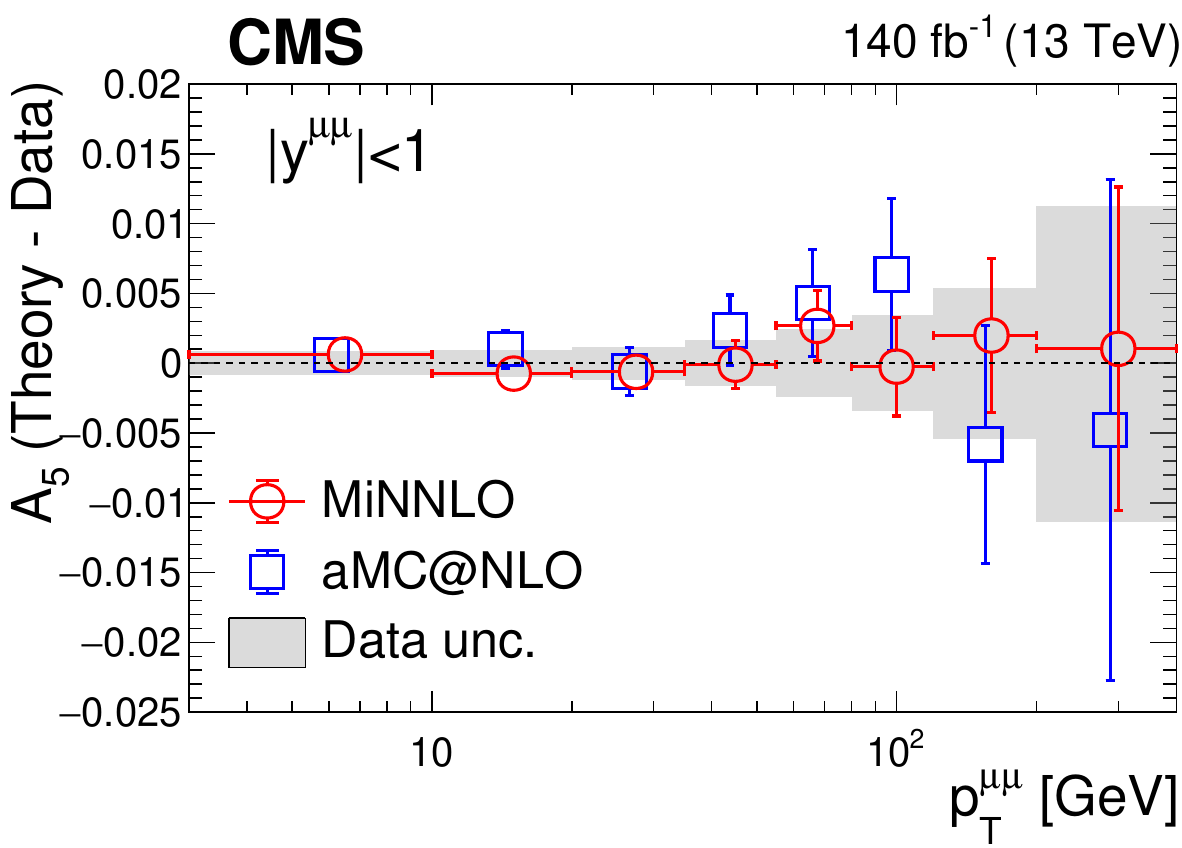}
\includegraphics[width=0.42\textwidth]{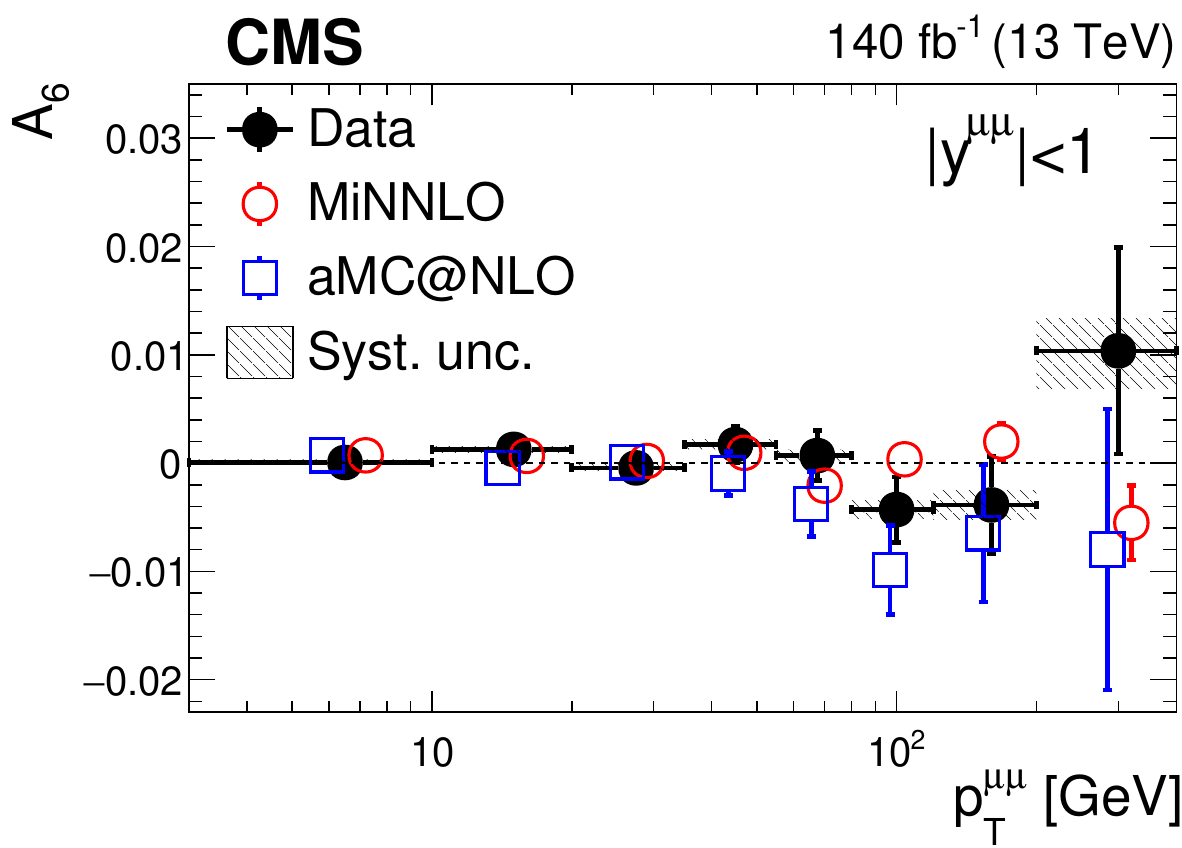}
\includegraphics[width=0.42\textwidth]{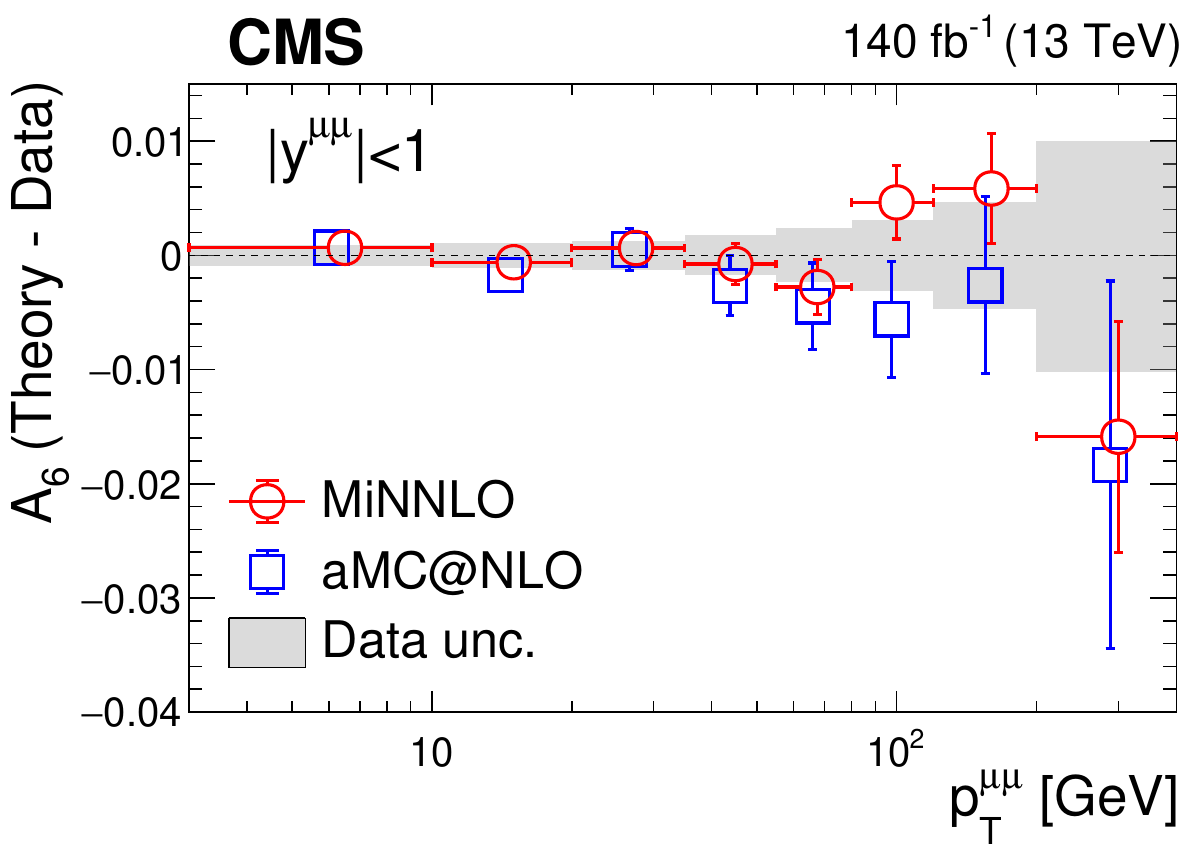}
\includegraphics[width=0.42\textwidth]{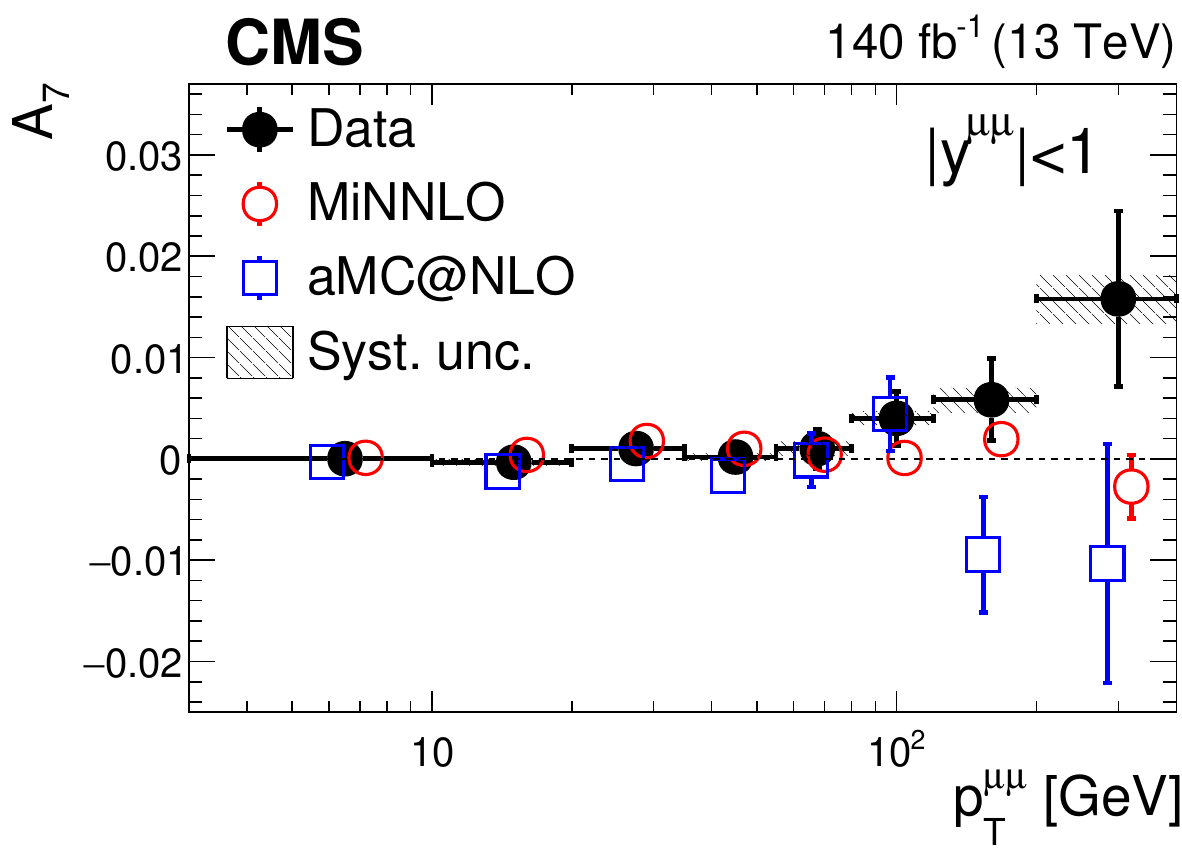}
\includegraphics[width=0.42\textwidth]{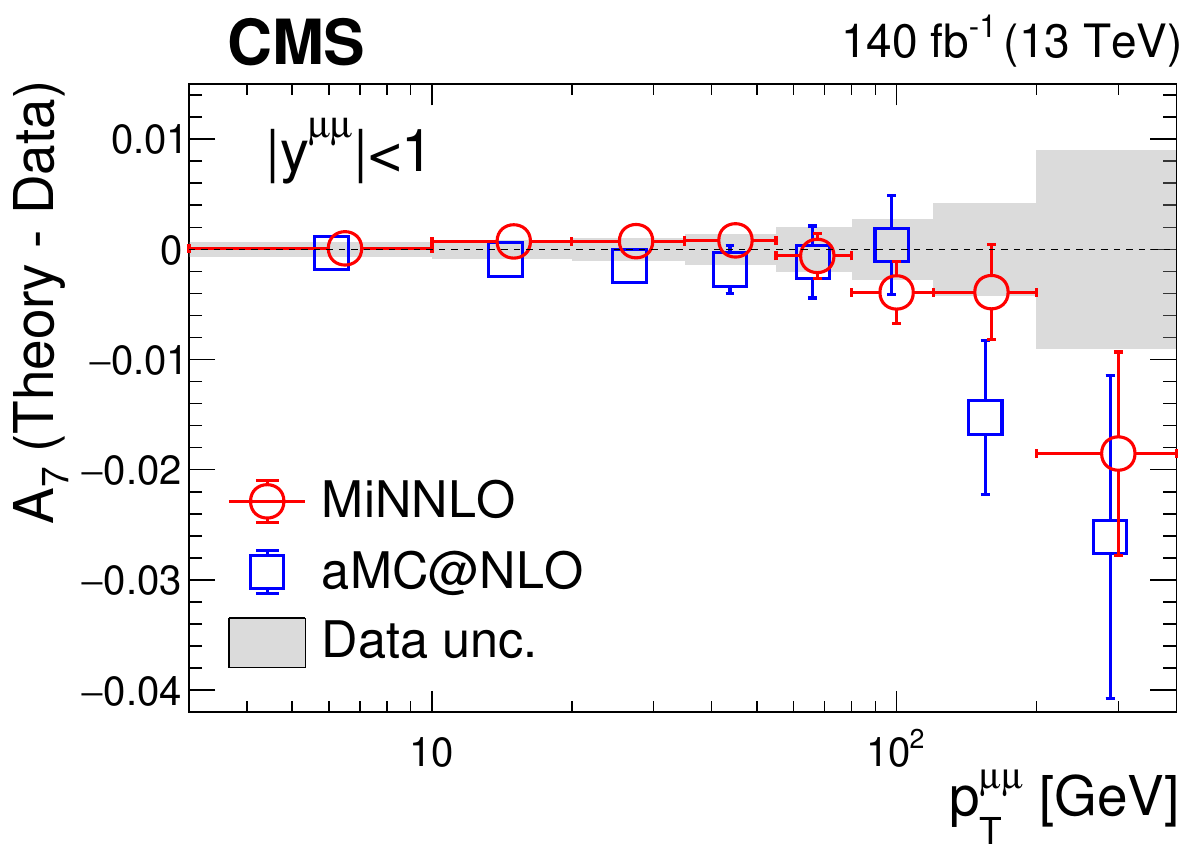}
\caption{Same as Fig.~\ref{CoeffsWithUncert_A0A3_Y0}, for the polarization coefficients $A_4$ to $A_7$.}
\label{CoeffsWithUncert_A4A7_Y0}
\end{figure*}

\begin{figure*}[htbp!]
\centering 
\includegraphics[width=0.42\textwidth]{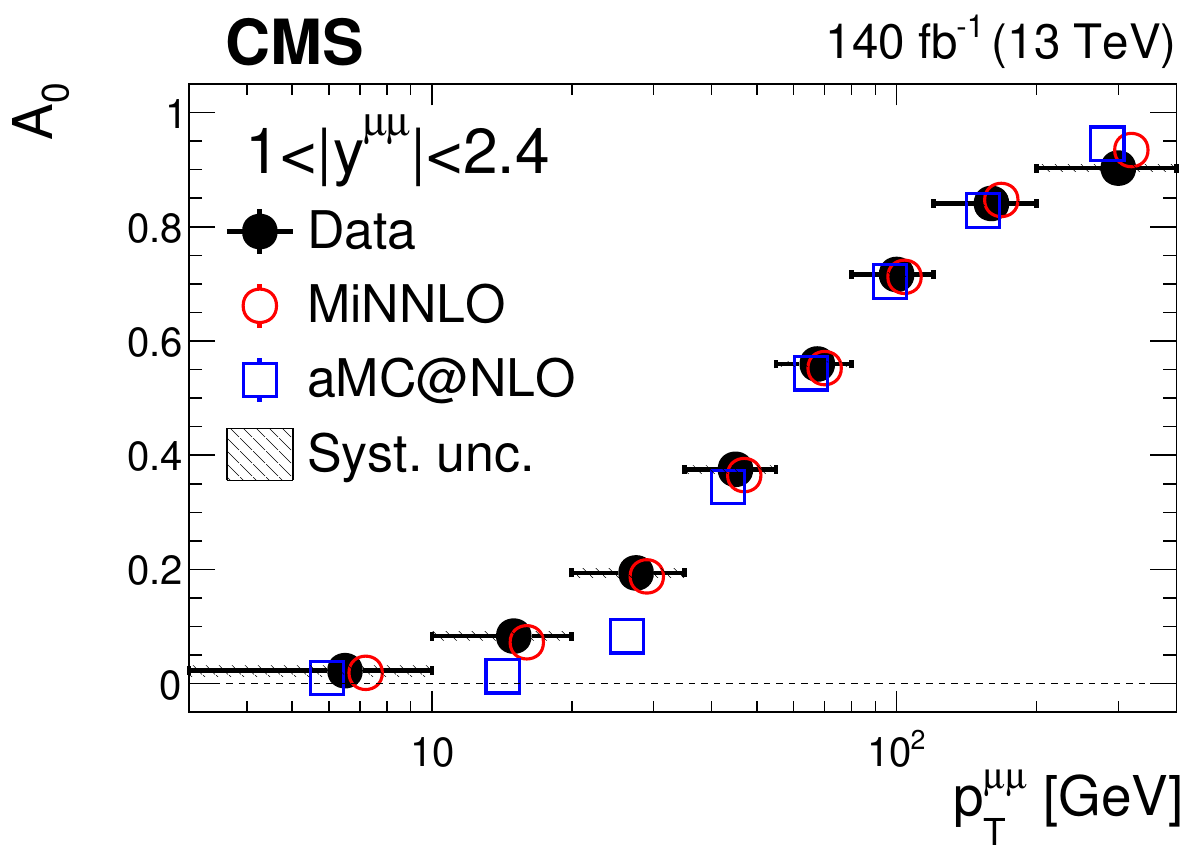}
\includegraphics[width=0.42\textwidth]{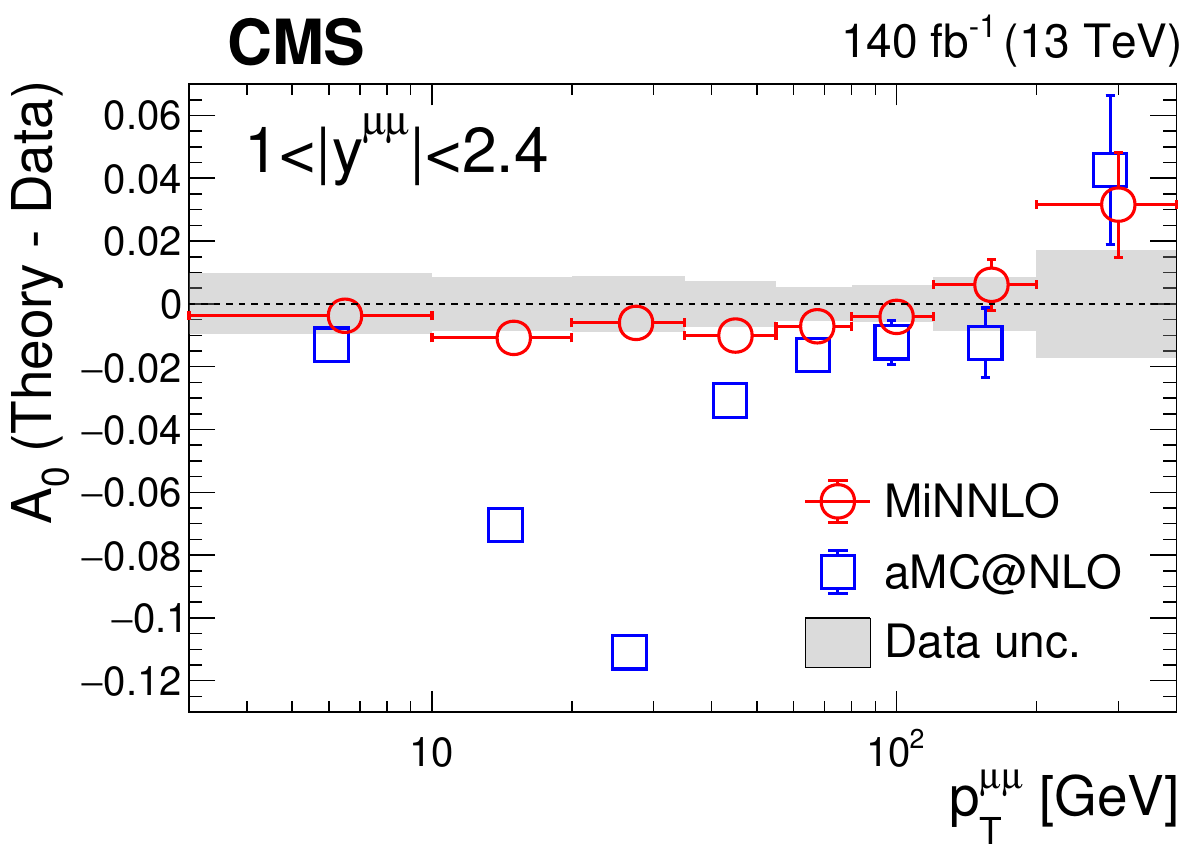}
\includegraphics[width=0.42\textwidth]{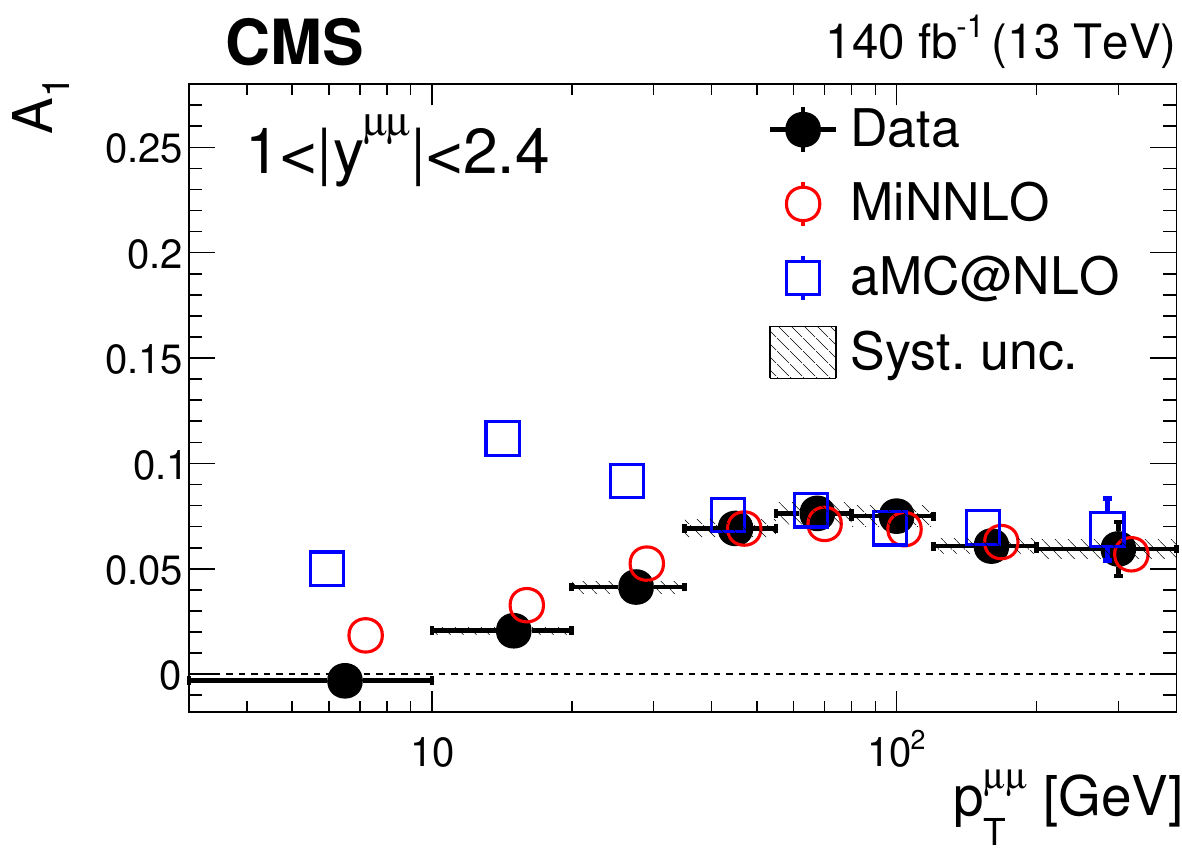}
\includegraphics[width=0.42\textwidth]{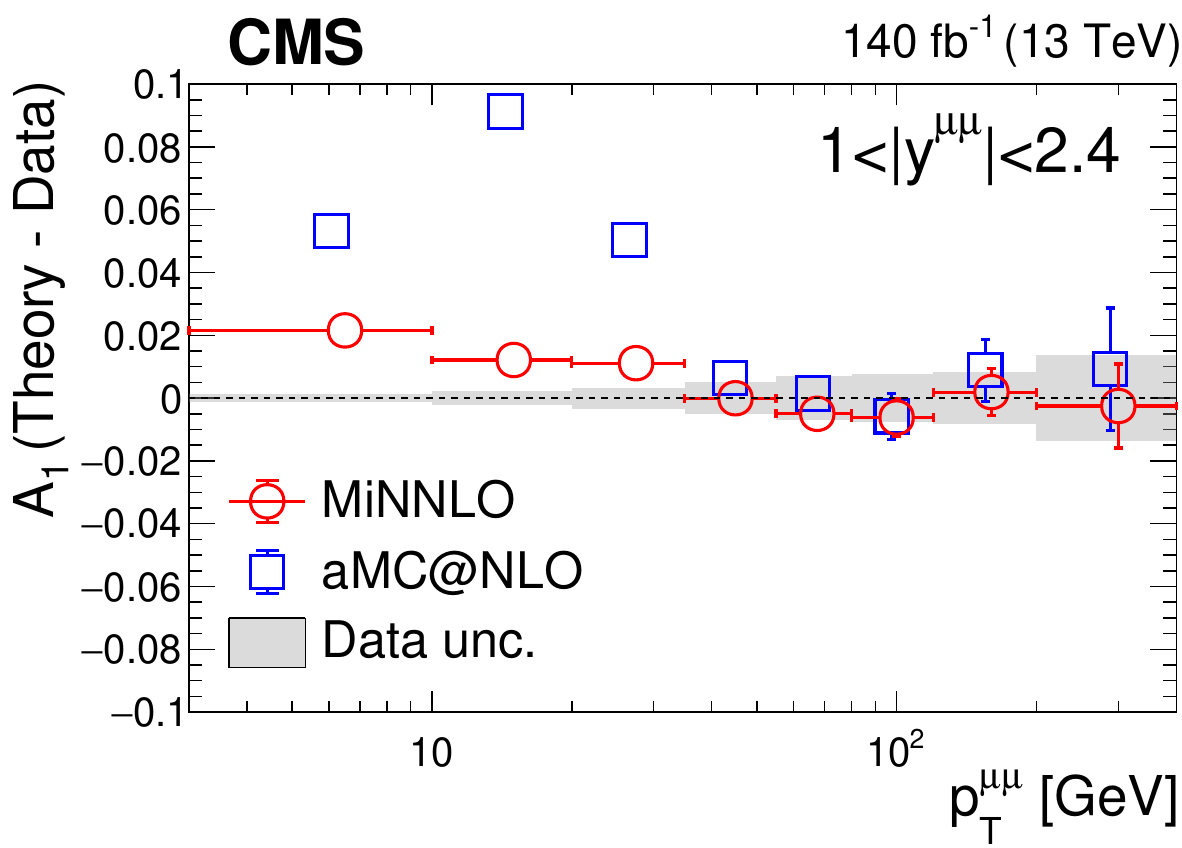}
\includegraphics[width=0.42\textwidth]{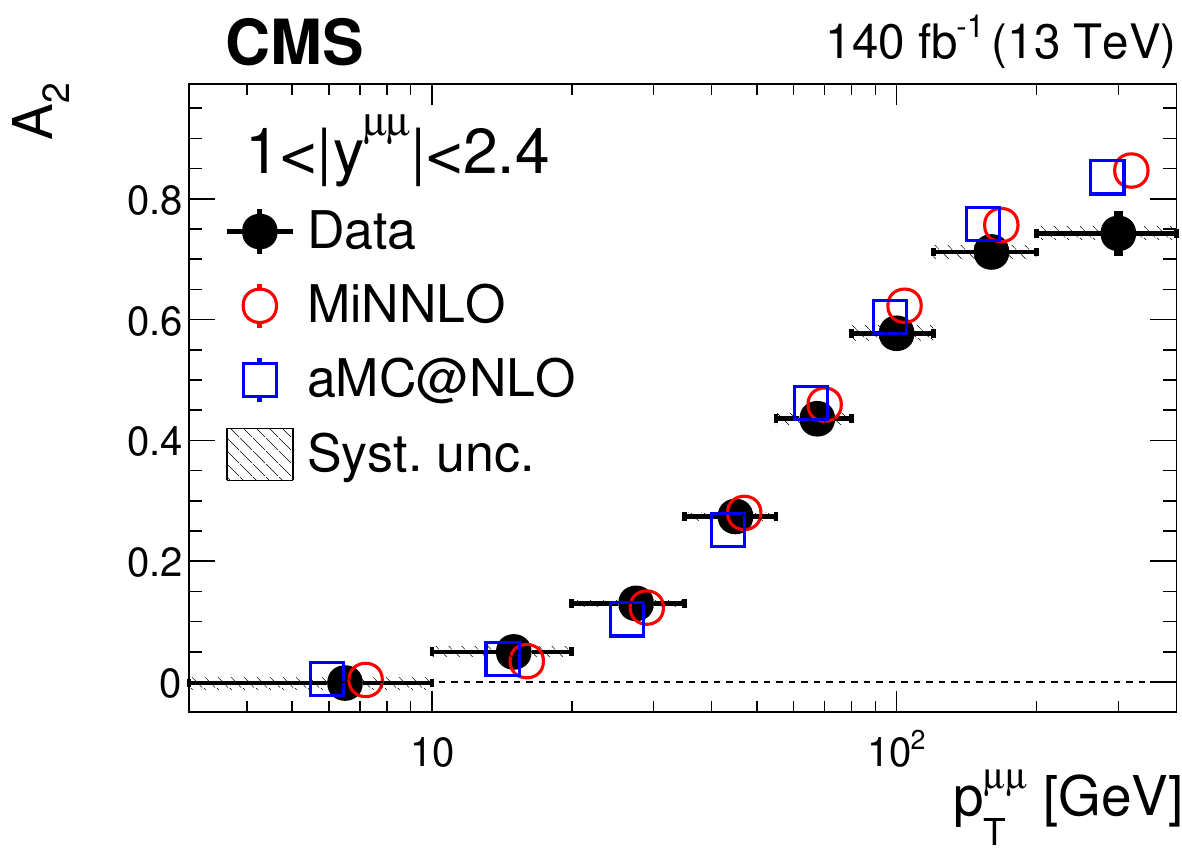}
\includegraphics[width=0.42\textwidth]{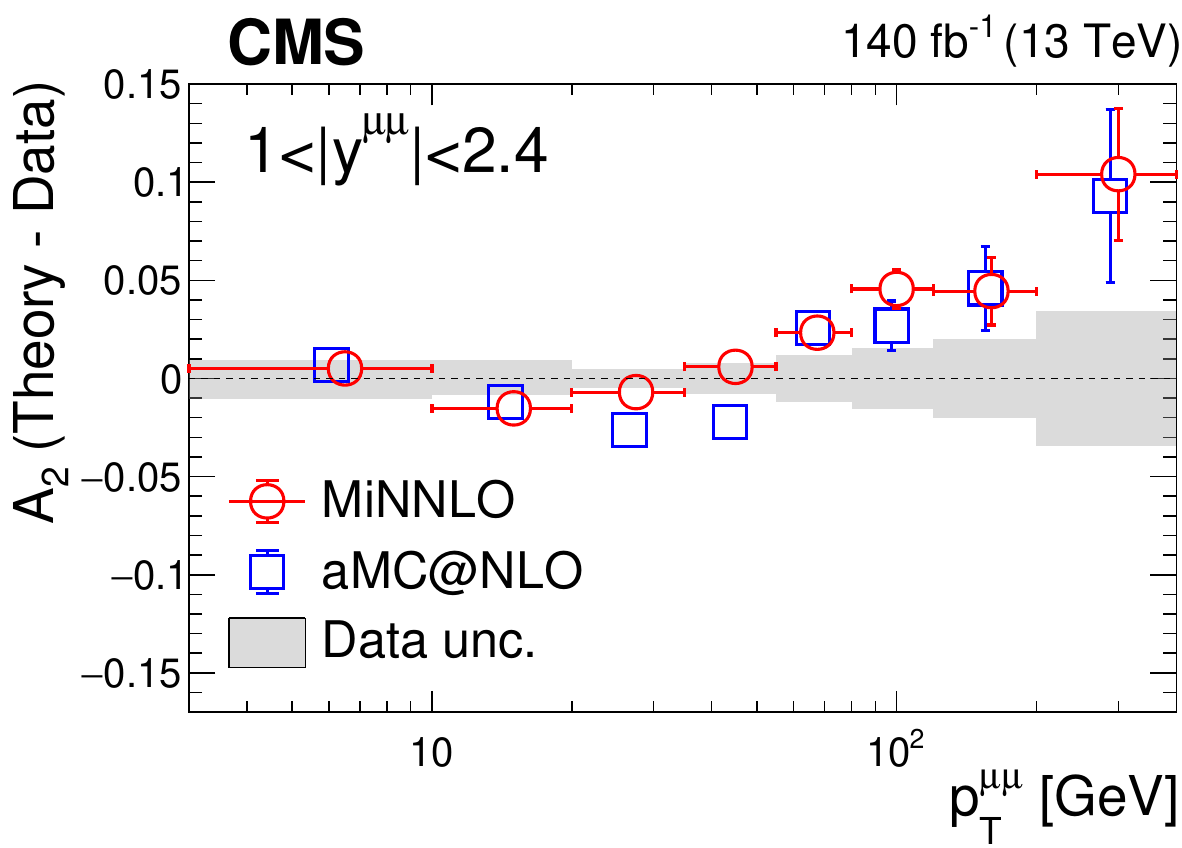}
\includegraphics[width=0.42\textwidth]{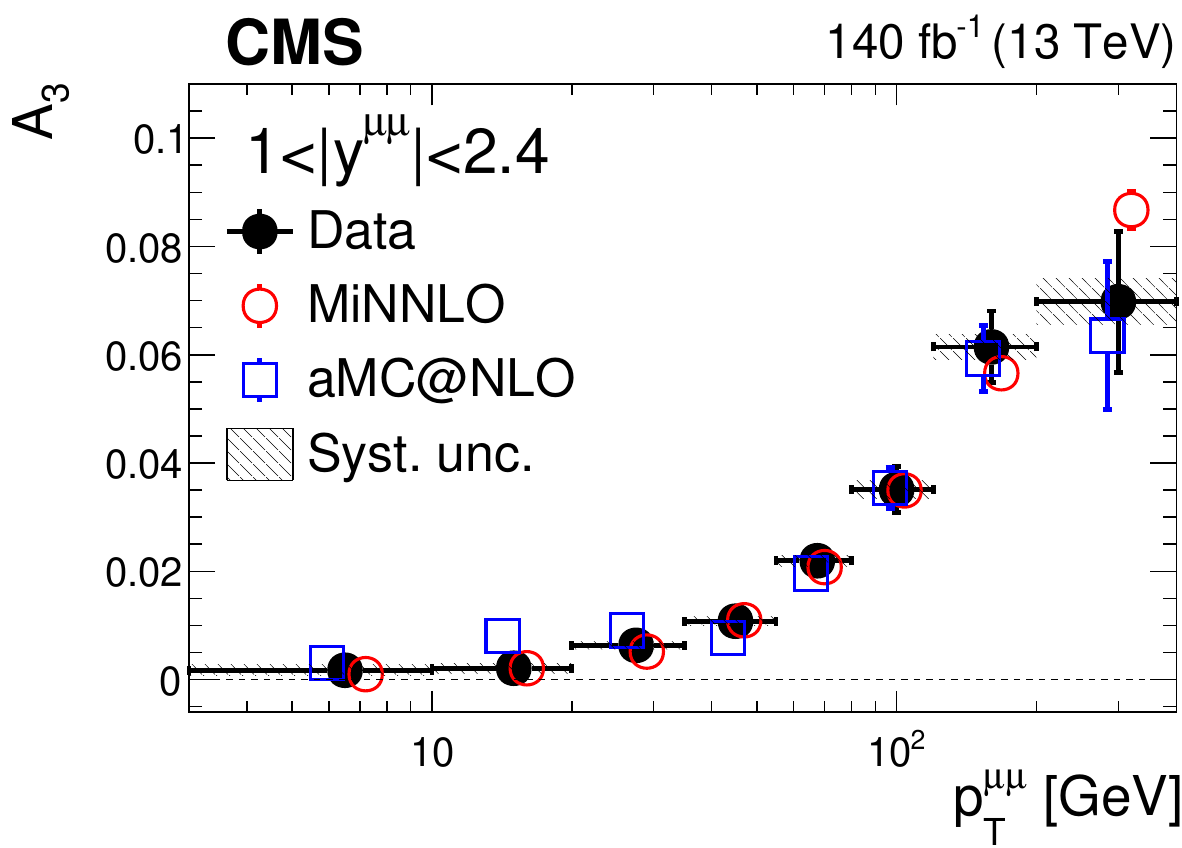}
\includegraphics[width=0.42\textwidth]{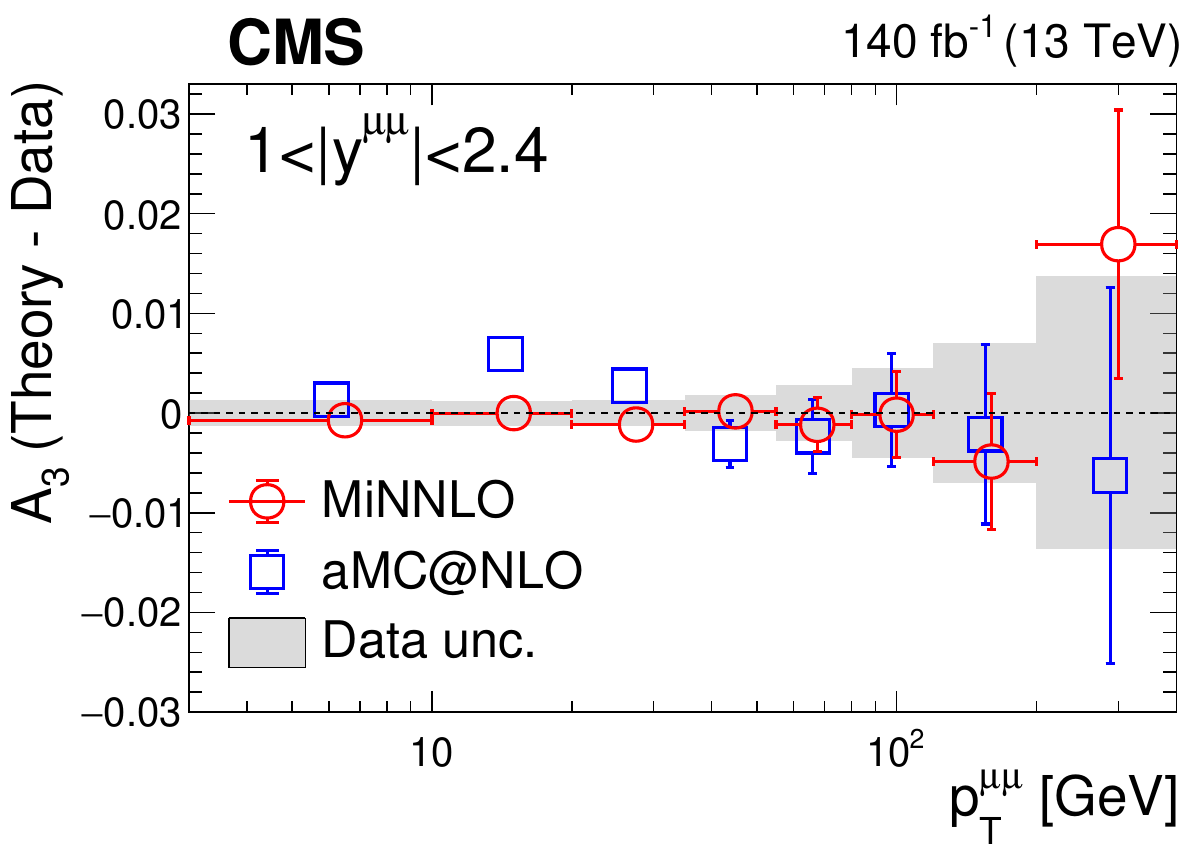}
\caption{Same as Fig.~\ref{CoeffsWithUncert_A0A3_Y0}, for the $1<\abs{\ymm}<2.4$ bin.}
\end{figure*}

\begin{figure*}[htbp!]
\centering 
\includegraphics[width=0.42\textwidth]{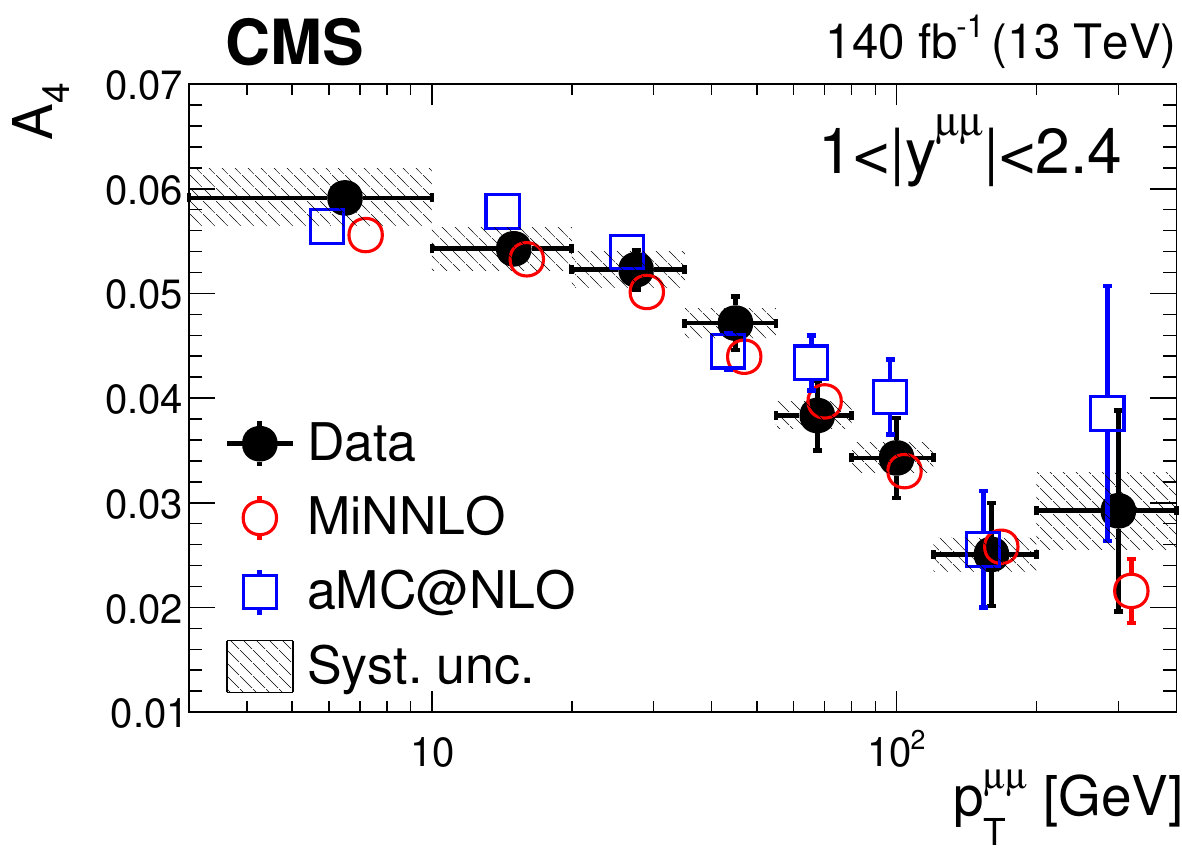}
\includegraphics[width=0.42\textwidth]{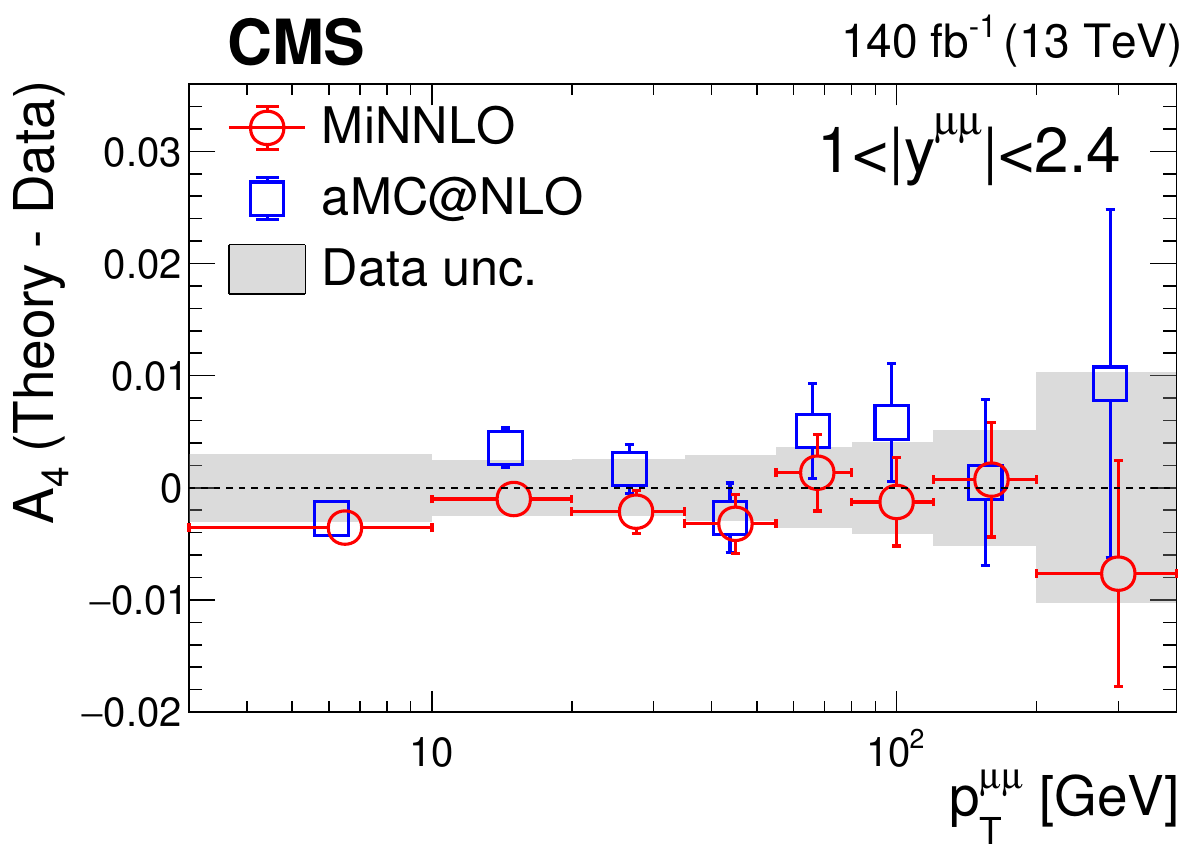}
\includegraphics[width=0.42\textwidth]{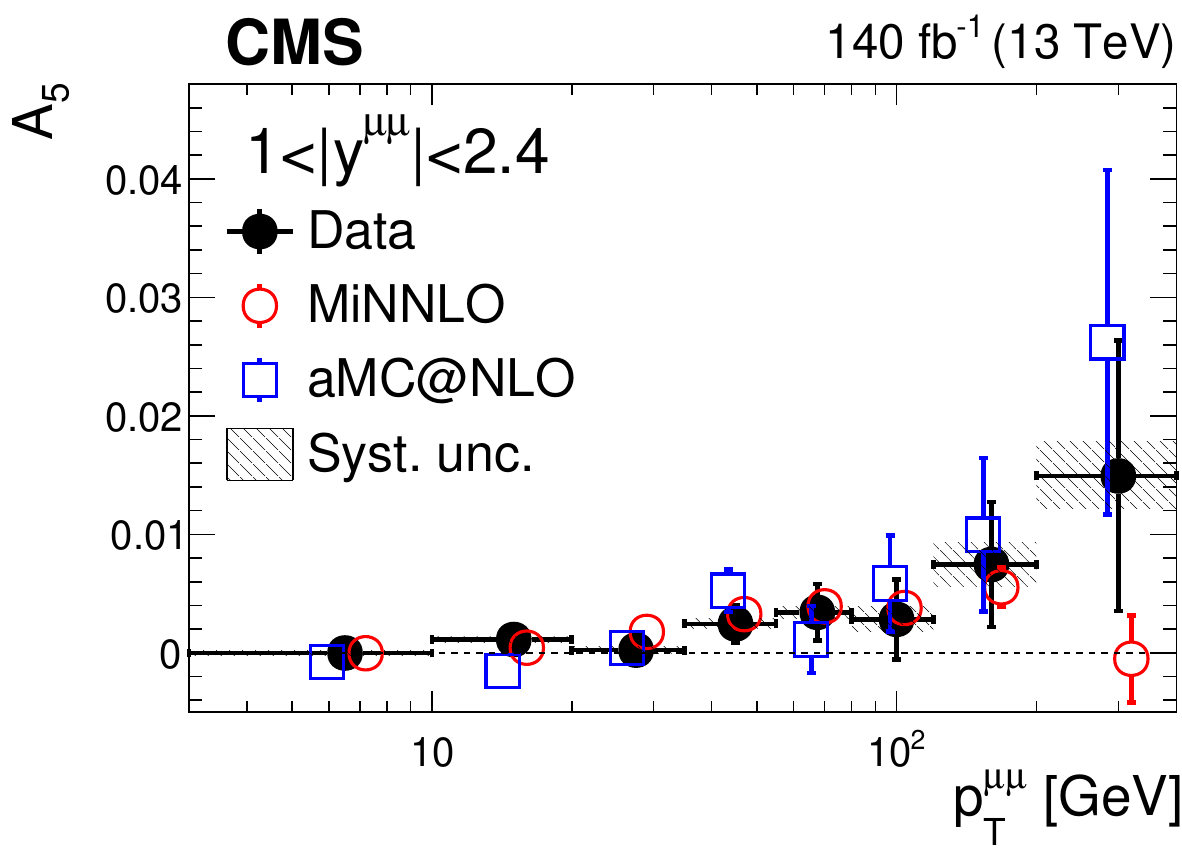}
\includegraphics[width=0.42\textwidth]{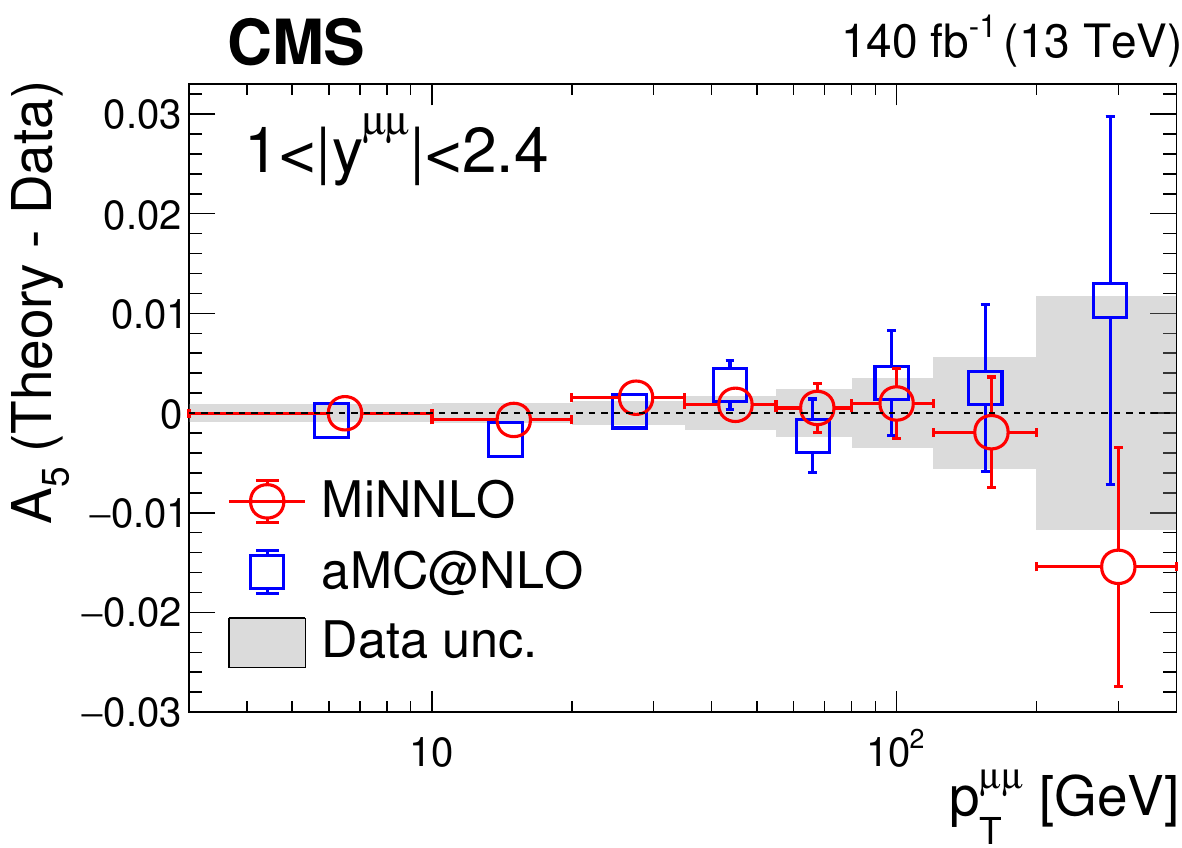}
\includegraphics[width=0.42\textwidth]{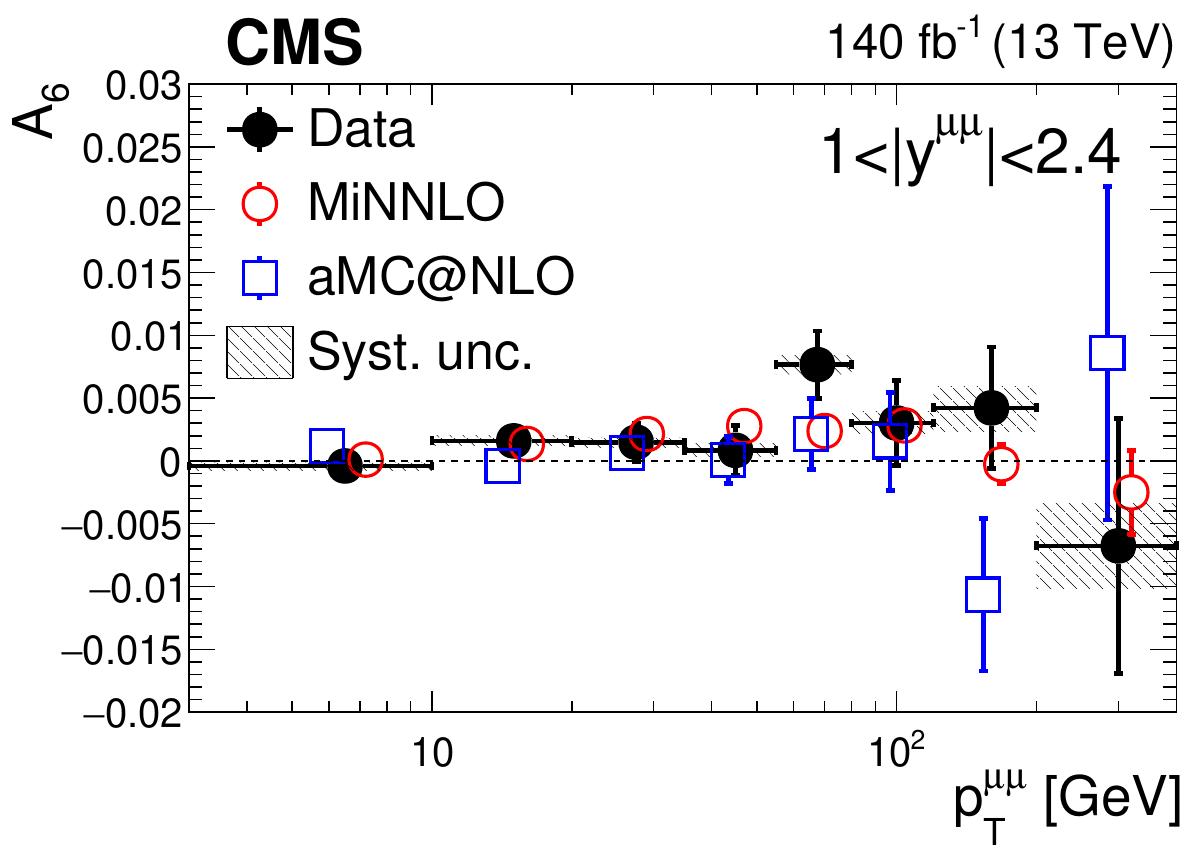}
\includegraphics[width=0.42\textwidth]{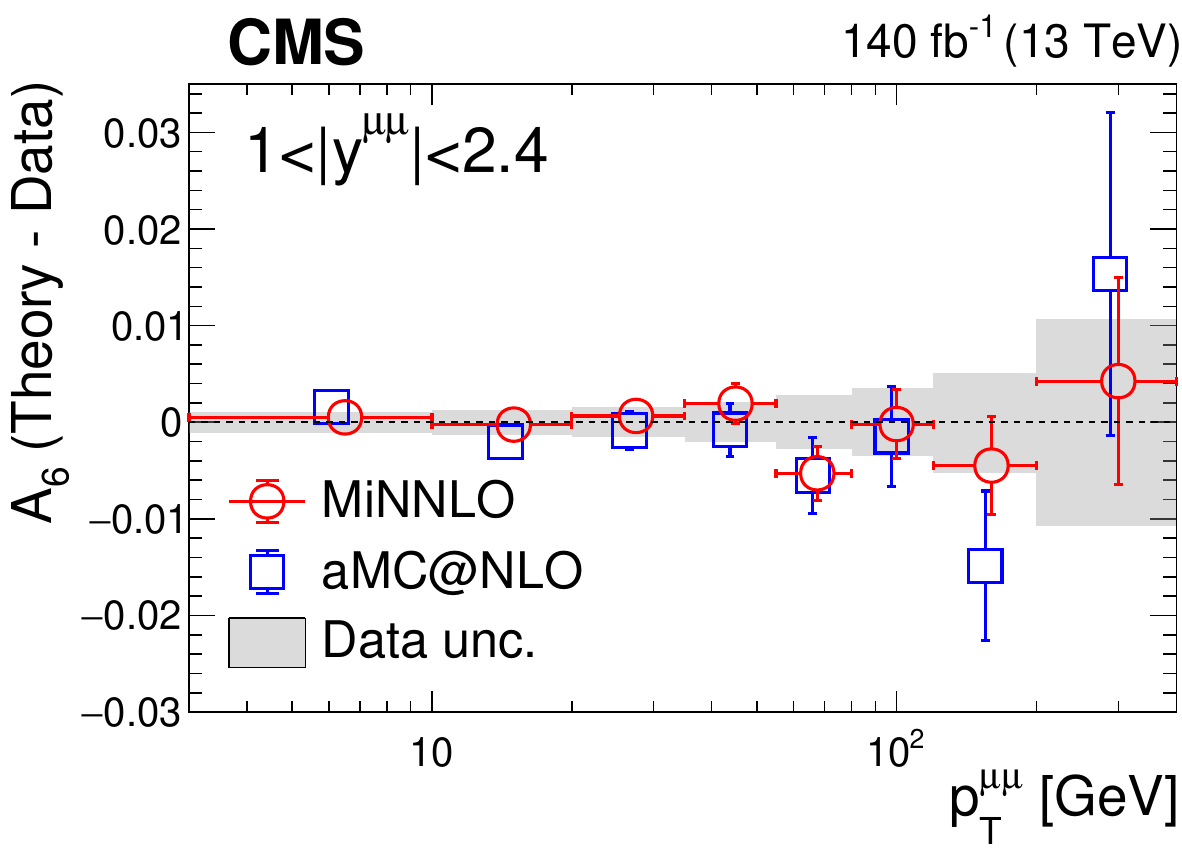}
\includegraphics[width=0.42\textwidth]{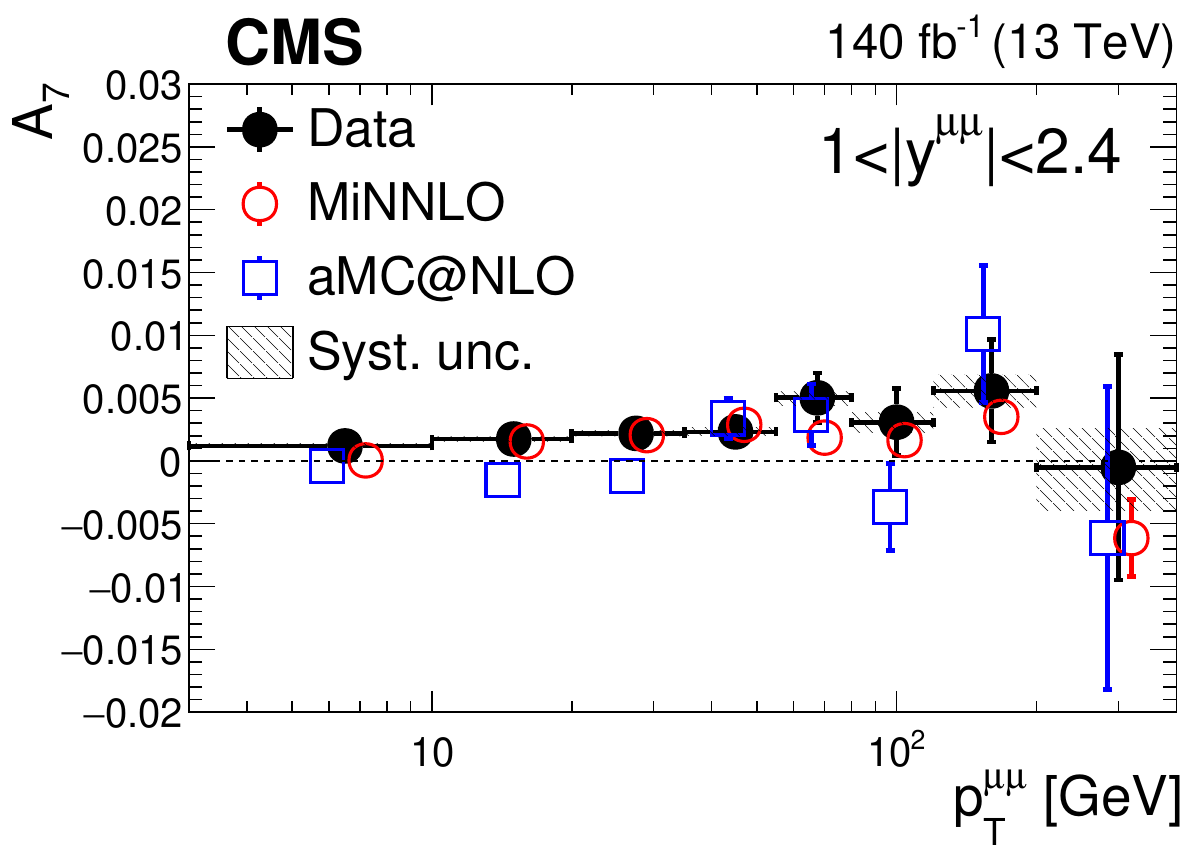}
\includegraphics[width=0.42\textwidth]{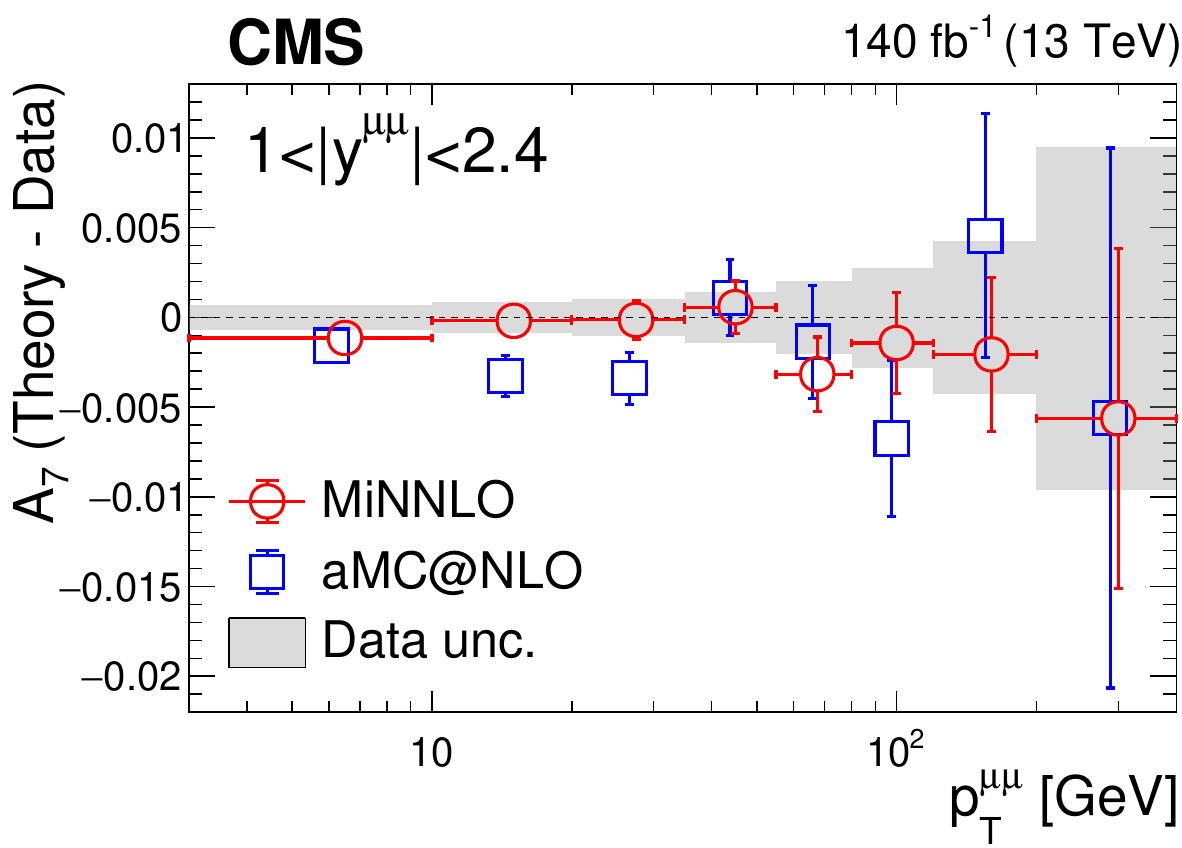}
\caption{Same as Fig.~\ref{CoeffsWithUncert_A4A7_Y0}, for the $1<\abs{\ymm}<2.4$ bin.}
\label{CoeffsWithUncert_Y1}
\end{figure*}

\begin{figure*}[ht]
\centering
\includegraphics[width=0.45\textwidth]{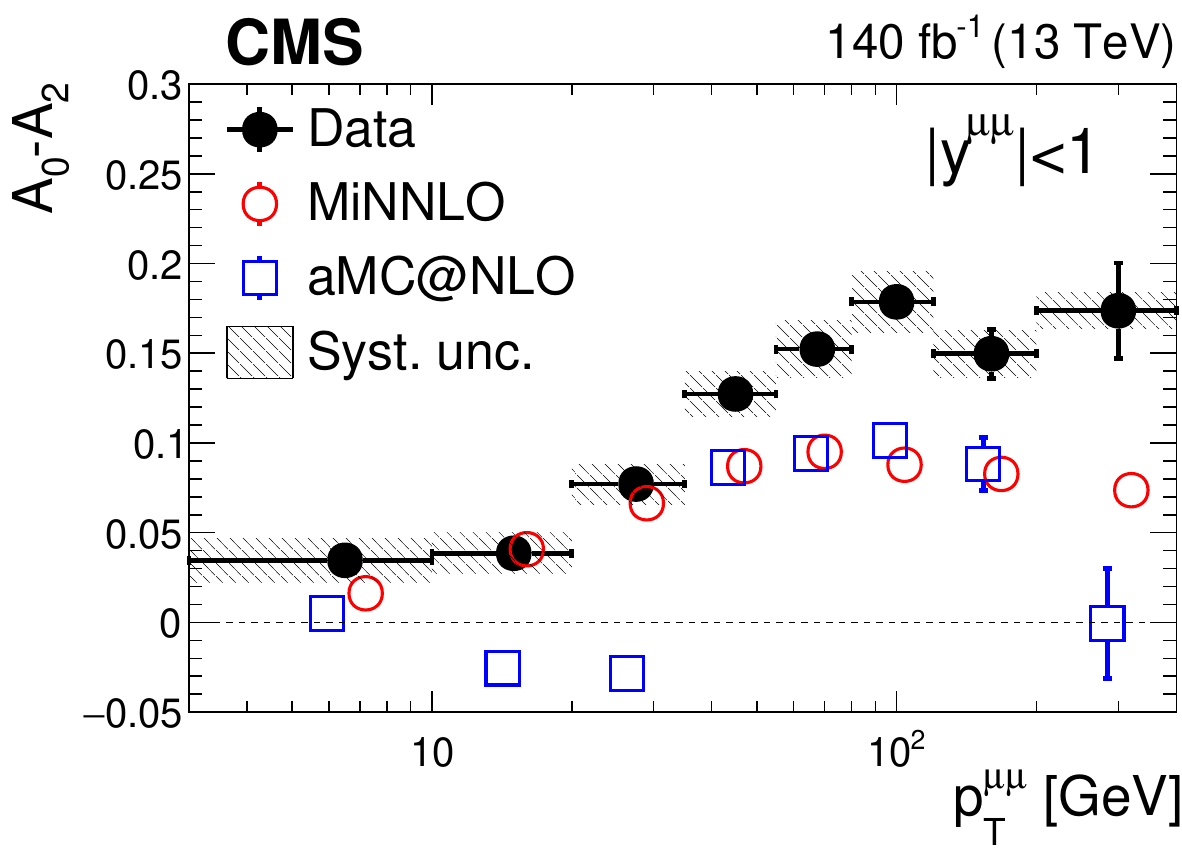}
\includegraphics[width=0.45\textwidth]{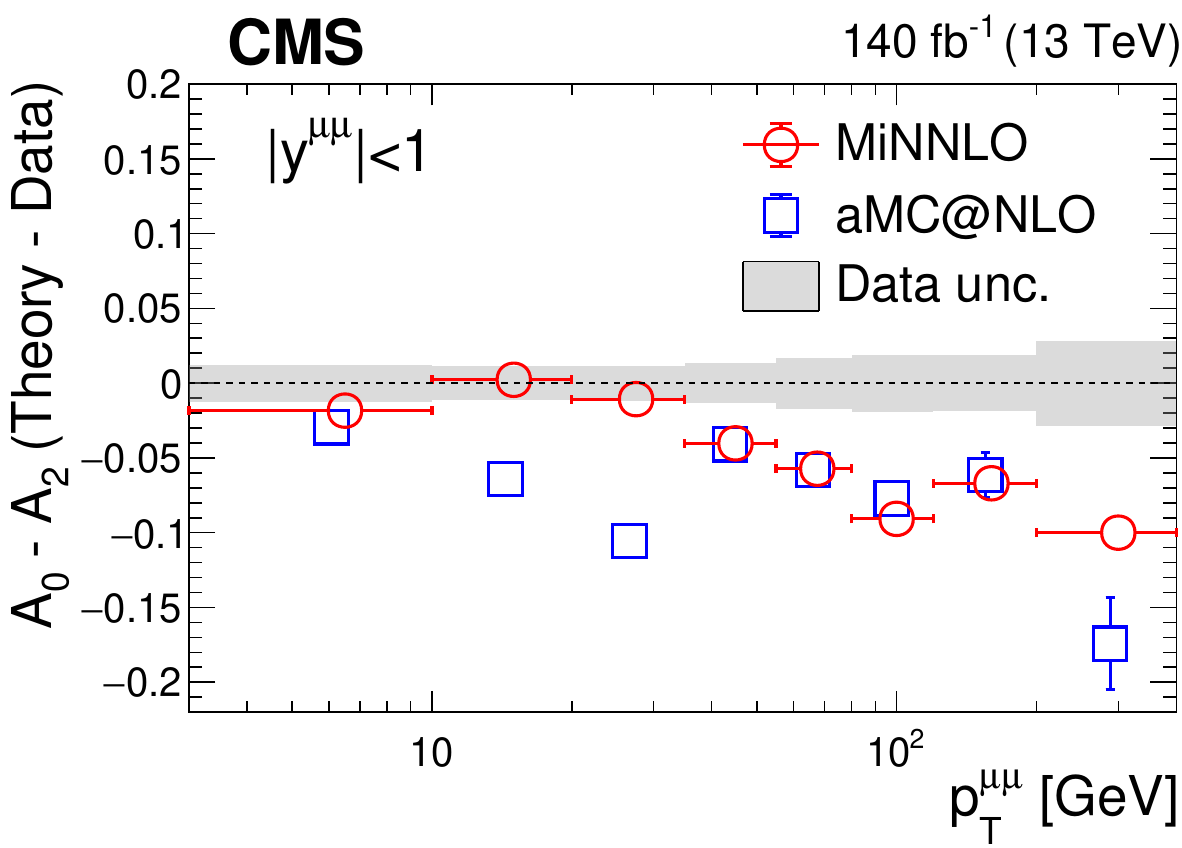}
\includegraphics[width=0.45\textwidth]{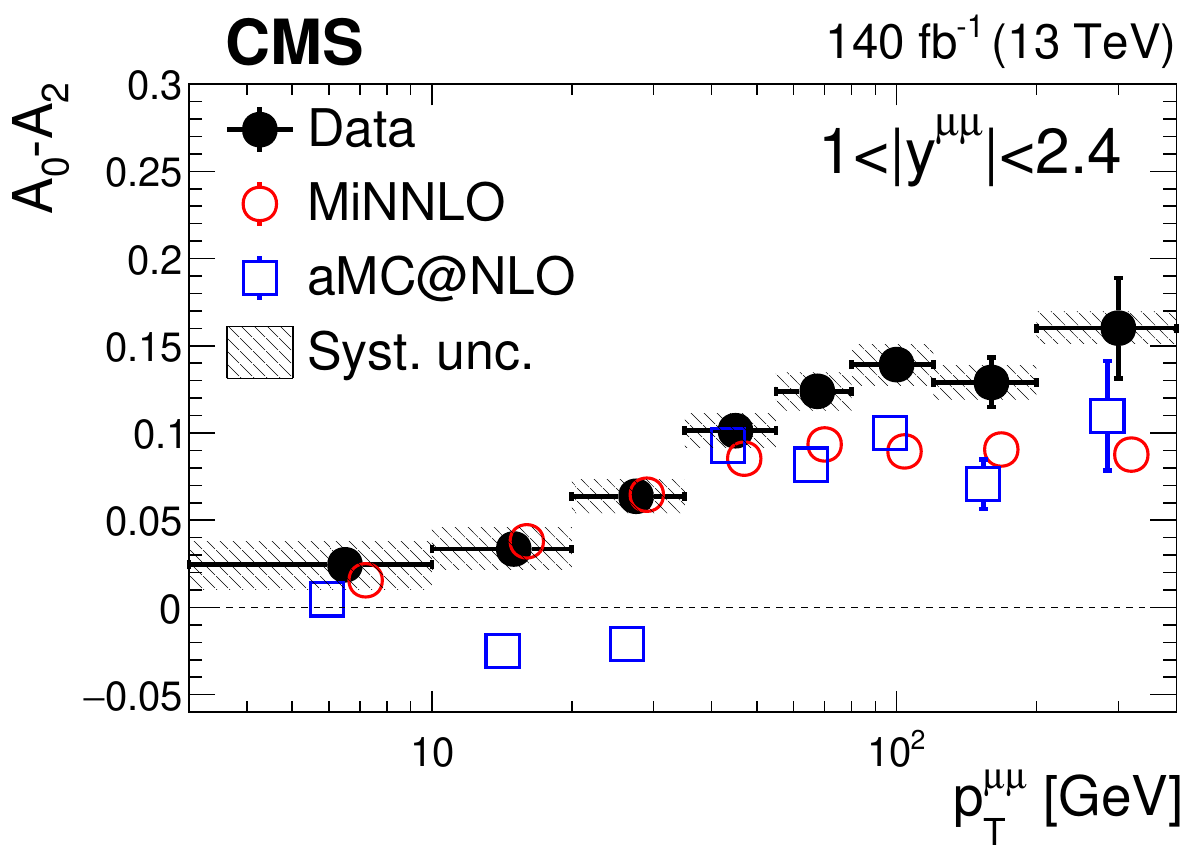}
\includegraphics[width=0.45\textwidth]{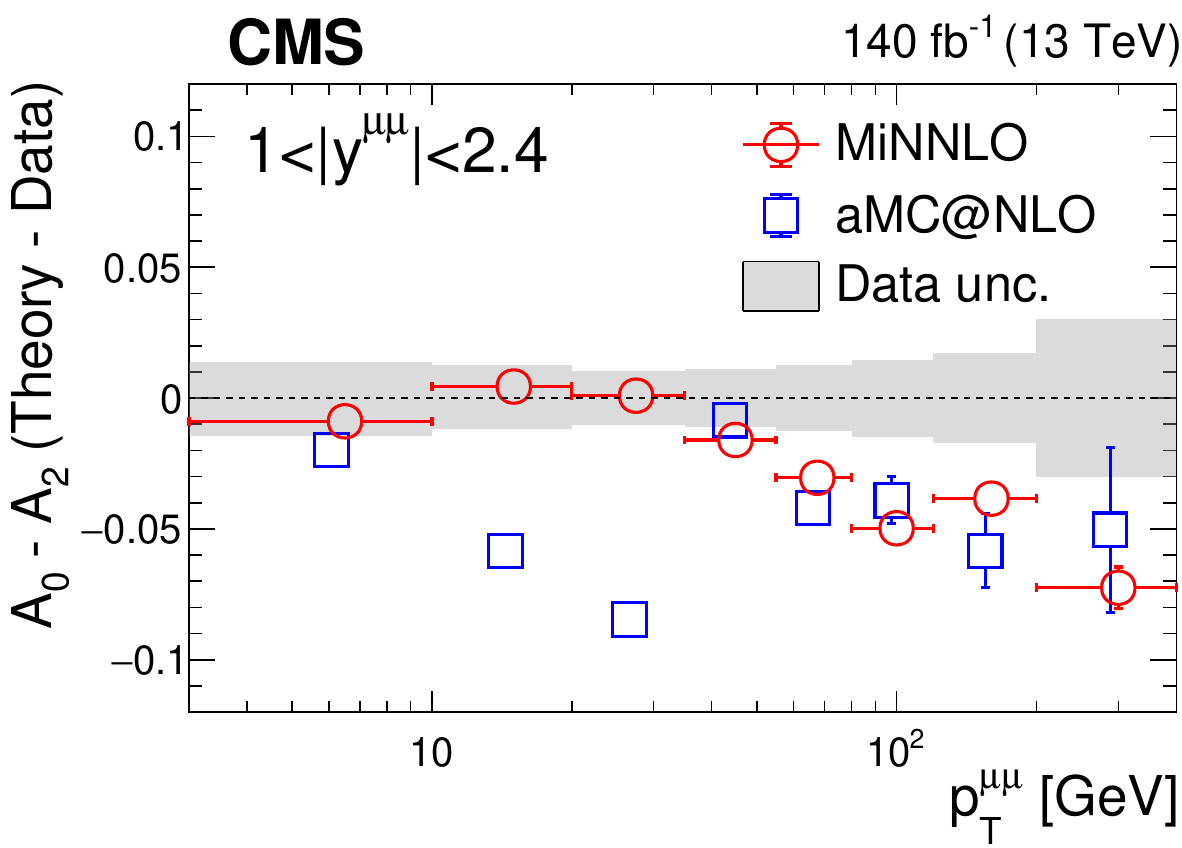}
\caption{Left: Difference $A_0-A_2$ measured in the CS frame in bins of \ptmm for $\abs{\ymm}<1$ (upper) and $1<\abs{\ymm}<2.4$ (lower).
Right: Corresponding differences between the predicted and measured values.}
\label{LTWithUncert}
\end{figure*}

Figures~\ref{CoeffsWithUncert_A0A3_Y0}--\ref{CoeffsWithUncert_Y1} present the measured $A_0$ to $A_7$ angular coefficients, in the CS frame, for dimuons with invariant mass $81<\mmm<101$\GeV, in two $\abs{\ymm}$ bins, while Fig.~\ref{LTWithUncert} presents the corresponding $A_0-A_2$ values. The \POWHEG\!+\textsc{MiNNLO} and \MGvATNLO predictions are also shown in the figures. The measured values are provided in Tables~\ref{TableAi1}--\ref{TableLT} and in the HEPData record of this analysis~\cite{HEPData}.

The $A_0$ values from data and \POWHEG\!+\textsc{MiNNLO} simulation agree within systematic and statistical uncertainties for low- and high-\ptmm bins, whereas in the intermediate \ptmm range we observe a discrepancy of about three standard deviations. The \MGvATNLO predictions significantly underestimate $A_0$ in the range $10 < \ptmm < 35$\GeV. The measured $A_2$ values coincide with the predicted \POWHEG\!+\textsc{MiNNLO} and \MGvATNLO values at low \ptmm, but are slightly overestimated by the MC for $\ptmm > 50$\GeV. In the 10--35\GeV \ptmm range, the \POWHEG\!+\textsc{MiNNLO} predictions are in good agreement with the measured $A_0 - A_2$ values, while the \MGvATNLO predictions show a significant discrepancy. For $\ptmm > 35$\GeV, instead, both MC predictions equally underestimate the measurement. The distribution of $A_1$ is significantly overestimated by \POWHEG\!+\textsc{MiNNLO} for $\ptmm < 35$\GeV in the $1 < \abs{\ymm} < 2.4$ region, and also overestimated by \MGvATNLO predictions in both rapidity regions at low \ptmm. The coefficients $A_5$, $A_6$, and $A_7$ have values of the order of $10^{-3}$ and are consistent with zero within uncertainties in most of the bins in data and both MC simulations. The most significant deviation from zero of the measured values for these coefficients, across all bins, is at 2.8 standard deviations for $A_6$ in the bin $55 < \ptmm < 80$\GeV, $1.2 < \abs{\ymm} < 2.4$. In general, the agreement between data and both MC simulations is good for the coefficients $A_3$--$A_7$. Nevertheless, in most cases the \POWHEG\!+\textsc{MiNNLO} predictions are closer to the data, which likely indicates the importance of including higher-order QCD corrections. The observed discrepancy between the measurement and the \MGvATNLO prediction for the $A_0$ and $A_1$ coefficients at $\ptmm<35$\GeV may be associated with the choice of the jet merging scale at 30\GeV, which affects the matching between matrix-element and parton shower emissions in the low-\ptmm region and with the choice of the parton shower simulation. A more detailed study is required to establish its origin.

The systematic uncertainty is dominant for coefficients $A_0$ and $A_1$ for $\ptmm < 55$\GeV, and for coefficients $A_2$, $A_3$, and $A_4$ for $\ptmm < 200$\GeV. For the Lam--Tung relation, the systematic uncertainty is larger than the statistical in all \ptmm bins except the highest two. For the other cases, the statistical uncertainty dominates.

The angular polarization coefficients measured at $\sqrt{s}=13$\TeV are also consistent with the values measured by the ATLAS and CMS Collaborations at $\sqrt{s}=8$\TeV~\cite{ ATLAS_AiMeasur_8TeV_2016, CMS_AiMeasur_8TeV_2015}. The sixfold larger integrated luminosity of the data and the higher production cross section has reduced the statistical uncertainty significantly. The deviations of the $A_6$ coefficient from zero, observed by ATLAS in the $1 < \abs{\ymm} < 2.4$ range at $\ptmm > 50$\GeV, are also visible at $\sqrt{s}=13$\TeV, but their magnitudes are noticeably smaller.

\section{Summary}

The first measurement of the full set of angular polarization coefficients $A_0$--$A_7$ in the Drell--Yan dimuon channel in the central rapidity range $0<\abs{\ymm}<2.4$ and $\ptmm<400$\GeV at $\sqrt{s}=13\TeV$ is presented. The coefficients were determined double differentially in bins of transverse momentum and rapidity of the dimuon in the 81--101\GeV invariant mass range. The results are compared with state-of-the-art theoretical predictions at next-to-next-to-leading order in QCD and show consistency within uncertainties in most of the phase space.

The presented results provide a comprehensive characterization of the angular structure of dilepton production and offer stringent constraints on theoretical descriptions of electroweak vector boson production mechanisms and underlying partonic dynamics~\cite{Zhang_EtractBoerMulder_2008, TMDandAC_2024}. The achieved precision and multidimensional binning establish a valuable benchmark for future phenomenological studies and for testing higher-order calculations~\cite{HAGIWARA_ToddAsym_1983, HAGIWARA_ProbOneLoop_1992}.

The most significant uncertainties vary across different coefficients and kinematic regions. The uncertainties on $A_0, A_1, A_2$ are mostly sensitive to the efficiency correction uncertainty while those on $A_3$ and $A_4$ also depend on the PDF$+\alpS(m_{\PZ})$ uncertainty. The statistics of simulated samples contribute to the uncertainty of all coefficients in the kinematic range $\ptmm>80$ \GeV.

\begin{acknowledgments}
We congratulate our colleagues in the CERN accelerator departments for the excellent performance of the LHC and thank the technical and administrative staffs at CERN and at other CMS institutes for their contributions to the success of the CMS effort. In addition, we gratefully acknowledge the computing centers and personnel of the Worldwide LHC Computing Grid and other centers for delivering so effectively the computing infrastructure essential to our analyses. Finally, we acknowledge the enduring support for the construction and operation of the LHC, the CMS detector, and the supporting computing infrastructure provided by the following funding agencies: SC (Armenia), BMBWF and FWF (Austria); FNRS and FWO (Belgium); CNPq, CAPES, FAPERJ, FAPERGS, and FAPESP (Brazil); MES and BNSF (Bulgaria); CERN; CAS, MoST, and NSFC (China); MINCIENCIAS (Colombia); MSES and CSF (Croatia); RIF (Cyprus); SENESCYT (Ecuador); ERC PRG and PSG, TARISTU24-TK10 and MoER TK202 (Estonia); Academy of Finland, MEC, and HIP (Finland); CEA and CNRS/IN2P3 (France); SRNSF (Georgia); BMFTR, DFG, and HGF (Germany); GSRI (Greece); MATE and NKFIH (Hungary); DAE and DST (India); IPM (Iran); SFI (Ireland); INFN (Italy); MSIT and NRF (Republic of Korea); MES (Latvia); LMTLT (Lithuania); MOE and UM (Malaysia); BUAP, CINVESTAV, CONACYT, LNS, SEP, and UASLP-FAI (Mexico); MOS (Montenegro); MBIE (New Zealand); PAEC (Pakistan); MSHE, NSC, and NAWA (Poland); FCT (Portugal); MESTD (Serbia); MICIU/AEI and PCTI (Spain); MOSTR (Sri Lanka); Swiss Funding Agencies (Switzerland); MST (Taipei); MHESI (Thailand); TUBITAK and TENMAK (T\"{u}rkiye); NASU (Ukraine); STFC (United Kingdom); DOE and NSF (USA).

\hyphenation{Rachada-pisek} Individuals have received support from the Marie-Curie program and the European Research Council and Horizon 2020 Grant, contract Nos.\ 675440, 724704, 752730, 758316, 765710, 824093, 101115353, 101002207, 101001205, and COST Action CA16108 (European Union); the Leventis Foundation; the Alfred P.\ Sloan Foundation; the Alexander von Humboldt Foundation; the Science Committee, project no. 22rl-037 (Armenia); the Fonds pour la Formation \`a la Recherche dans l'Industrie et dans l'Agriculture (FRIA) and Fonds voor Wetenschappelijk Onderzoek contract No. 1228724N (Belgium); the Beijing Municipal Science \& Technology Commission, No. Z191100007219010, the Fundamental Research Funds for the Central Universities, the Ministry of Science and Technology of China under Grant No. 2023YFA1605804, the Natural Science Foundation of China under Grant No. 12535004, and USTC Research Funds of the Double First-Class Initiative No.\ YD2030002017 (China); the Ministry of Education, Youth and Sports (MEYS) of the Czech Republic; the Shota Rustaveli National Science Foundation, grant FR-22-985 (Georgia); the Deutsche Forschungsgemeinschaft (DFG), among others, under Germany's Excellence Strategy -- EXC 2121 ``Quantum Universe" -- 390833306, and under project number 400140256 - GRK2497; the Hellenic Foundation for Research and Innovation (HFRI), Project Number 2288 (Greece); the Hungarian Academy of Sciences, the New National Excellence Program - \'UNKP, the NKFIH research grants K 131991, K 138136, K 143460, K 143477, K 147557, K 146913, K 146914, K 147048, TKP2021-NKTA-64, and 2025-1.1.5-NEMZ\_KI-2025-00004, and MATE KKP and KKPCs Research Excellence and Flagship Research Groups grants (Hungary); the Council of Science and Industrial Research, India; ICSC -- National Research Center for High Performance Computing, Big Data and Quantum Computing, FAIR -- Future Artificial Intelligence Research, and CUP I53D23001070006 (Mission 4 Component 1), funded by the NextGenerationEU program, the Italian Ministry of University and Research (MUR) under Bando PRIN 2022 -- CUP I53C24002390006, PRIN PRIMULA 2022RBYK7T (Italy); the Latvian Council of Science; the Ministry of Science and Higher Education, project no. 2022/WK/14, and the National Science Center, contracts Opus 2021/41/B/ST2/01369, 2021/43/B/ST2/01552, 2023/49/B/ST2/03273, and the NAWA contract BPN/PPO/2021/1/00011 (Poland); the Funda\c{c}\~ao para a Ci\^encia e a Tecnologia (Portugal); the National Priorities Research Program by Qatar National Research Fund; MICIU/AEI/10.13039/501100011033, ERDF/EU, ``European Union NextGenerationEU/PRTR", projects PID2022-142604OB-C21, PID2022-139519OB-C21, PID2023-147706NB-I00, PID2023-148896NB-I00, PID2023-146983NB-I00, PID2023-147115NB-I00, PID2023-148418NB-C41, PID2023-148418NB-C42, PID2023-148418NB-C43, PID2023-148418NB-C44, PID2024-158190NB-C22, RYC2021-033305-I, RYC2024-048719-I, CNS2023-144781, CNS2024-154769 and Plan de Ciencia, Tecnolog{\'i}a e Innovaci{\'o}n de Asturias, Spain; the Chulalongkorn Academic into Its 2nd Century Project Advancement Project, the National Science, Research and Innovation Fund program IND\_FF\_68\_369\_2300\_097, and the Program Management Unit for Human Resources \& Institutional Development, Research and Innovation, grant B39G680009 (Thailand); the Eric \& Wendy Schmidt Fund for Strategic Innovation through the CERN Next Generation Triggers project under grant agreement number SIF-2023-004; the Kavli Foundation; the Nvidia Corporation; the SuperMicro Corporation; the Welch Foundation, contract C-1845; and the Weston Havens Foundation (USA).
\end{acknowledgments}\section*{Data availability} Release and preservation of data used by the CMS Collaboration as the basis for publications is guided by the  \href{https://doi.org/10.7483/OPENDATA.CMS.1BNU.8V1W}{CMS data preservation, re-use and open access policy}.
\bibliography{auto_generated}

\appendix
\numberwithin{table}{section}
\section{Tables of angular polarization coefficients}
\label{app:tables}

This appendix collects tables listing the angular polarization coefficients $A_0$--$A_7$, as well as the $A_0-A_2$ difference, measured in the CS frame, as a function of \ptmm and in two $\abs{\ymm}$ intervals. These numerical values correspond to the data points shown in Figs.~\ref{CoeffsWithUncert_A0A3_Y0}--\ref{LTWithUncert}.

\begin{table*}[ht]
\topcaption{Measured angular coefficients, in bins of \ptmm (in \GeV), for $\abs{\ymm} < 1$.}
\label{TableAi1}
\centering
\begin{tabular}{c cccccccc}
\ptmm & 0--10 & 10--20 & 20--35 & 35--55 & 55--80 & 80--120 & 120--200 & 200--400 \\
\hline
$A_0$ & 0.030 & 0.086 & 0.205 & 0.392 & 0.572 & 0.723 & 0.848 & 0.922 \\
Stat. & $\pm$0.001 & $\pm$0.002 & $\pm$0.002 & $\pm$0.003 & $\pm$0.003 & $\pm$0.004 & $\pm$0.008 & $\pm$0.015 \\
Syst. & $\pm$0.012 & $\pm$0.011 & $\pm$0.010 & $\pm$0.006 & $\pm$0.002 
& \begin{tabular}{@{}c@{}}$+0.003$ \\[-1mm] $-0.004$\end{tabular} 
& $\pm$0.005 & $\pm$0.006 \\
\hline
$A_1$ & 0.0020 & 0.010 & 0.019 & 0.027 & 0.023 & 0.014 & 0.014 & $<$0.0001 \\
Stat. & $\pm$0.0009 & $\pm$0.001 & $\pm$0.002 & $\pm$0.002 & $\pm$0.004 & $\pm$0.005 & $\pm$0.006 & $\pm$0.0113 \\
Syst. & $\pm$0.0007 & $\pm$0.001 & $\pm$0.002 & $\pm$0.003 & $\pm$0.003 & $\pm$0.003 & $\pm$0.002 
& \begin{tabular}{@{}c@{}}$+0.0038$ \\[-1mm] $-0.0035$\end{tabular} \\
\hline
$A_2$ &$-0.004$ & 0.047 & 0.128 & 0.264 & 0.420 & 0.54 & 0.70 & 0.75 \\
Stat. & $\pm$0.002 & $\pm$0.002 & $\pm$0.002 & $\pm$0.003 & $\pm$0.006 & $\pm$0.01 & $\pm$0.02 & $\pm$0.03 \\
Syst. & $\pm$0.004 
& \begin{tabular}{@{}c@{}}$+0.004$ \\[-1mm] $-0.003$\end{tabular}
& $\pm$0.005 & $\pm$0.011 & $\pm$0.016 & $\pm$0.02 & $\pm$0.02 & $\pm$0.01 \\
\hline
$A_3$ & 0.0020 & 0.0011 & 0.0018 & 0.0032 & 0.008 & 0.011 & 0.025 & 0.037 \\
Stat. & $\pm$0.0007 & $\pm$0.0008 & $\pm$0.0011 & $\pm$0.0016 & $\pm$0.003 & $\pm$0.004 & $\pm$0.006 & $\pm$0.012 \\
Syst. & $\pm$0.0011 & $\pm$0.0009 & $\pm$0.0007 & $\pm$0.0008 & $\pm$0.001 
& \begin{tabular}{@{}c@{}}$+0.002$ \\[-1mm] $-0.001$\end{tabular}
& $\pm$0.002 & $\pm$0.004 \\
\hline
$A_4$ & 0.0125 & 0.011 & 0.0082 & 0.0098 & 0.0113 & 0.0080 & 0.001 & 0.008 \\
Stat. & $\pm$0.0009 & $\pm$0.001 & $\pm$0.0015 & $\pm$0.0020 & $\pm$0.0025 & $\pm$0.0031 & $\pm$0.004 & $\pm$0.009 \\
Syst. & $\pm$0.0015 & $\pm$0.001 & $\pm$0.0008 
& \begin{tabular}{@{}c@{}}$+0.0008$ \\[-1mm] $-0.0007$\end{tabular}
& $\pm$0.0008 & $\pm$0.0009 & $\pm$0.002 & $\pm$0.003 \\
\hline
$A_5$ &$ -0.0005$ & 0.0005 & 0.0005 & 0.0017 & $-0.0001$ & 0.0020 &$ -0.001 $& $-0.003$ \\
Stat. & $\pm$0.0008 & $\pm$0.0009 & $\pm$0.0012 & $\pm$0.0016 & $\pm$0.0024 & $\pm$0.0033 & $\pm$0.005 & $\pm$0.011 \\
Syst. & $\pm$0.0002 & $\pm$0.0002 & $\pm$0.0003 
& \begin{tabular}{@{}c@{}}$+0.0005$ \\[-1mm] $-0.0004$\end{tabular}
& $\pm$0.0005 & $\pm$0.0007 
& \begin{tabular}{@{}c@{}}$+0.001$ \\[-1mm] $-0.002$\end{tabular}
& $\pm$0.003 \\
\hline
$A_6$ & 0.0001 & 0.0013 & $-0.0004$ & 0.0017 & 0.0007 & $-0.0043$ & $-0.004$ & 0.010 \\
Stat. & $\pm$0.0009 & $\pm$0.0011 & $\pm$0.0013 & $\pm$0.0017 & $\pm$0.0023 & $\pm$0.0030 & $\pm$0.005 & $\pm$0.010 \\
Syst. & \begin{tabular}{@{}c@{}}$+0.0003$ \\[-1mm] $-0.0002$\end{tabular}
& $\pm$0.0003 & $\pm$0.0003 
& \begin{tabular}{@{}c@{}}$+0.0005$ \\[-1mm] $-0.0004$\end{tabular}
& \begin{tabular}{@{}c@{}}$+0.0006$ \\[-1mm] $-0.0005$\end{tabular}
& $\pm$0.0008 & $\pm$0.001 
& \begin{tabular}{@{}c@{}}$+0.003$ \\[-1mm] $-0.004$\end{tabular} \\
\hline
$A_7$ & $<$0.0001 & $-0.0003$ & 0.0010 & 0.0002 & 0.0010 & 0.0040 & 0.006 & 0.016 \\
Stat. & $\pm$0.0007 & $\pm$0.0008 & $\pm$0.0010 & $\pm$0.0014 & $\pm$0.0019 & $\pm$0.0027 & $\pm$0.004 & $\pm$0.009 \\
Syst. & $\pm$0.0002 
& \begin{tabular}{@{}c@{}}$+0.0002$ \\[-1mm] $-0.0003$\end{tabular}
& $\pm$0.0002 
& \begin{tabular}{@{}c@{}}$+0.0003$ \\[-1mm] $-0.0004$\end{tabular}
& \begin{tabular}{@{}c@{}}$+0.0007$ \\[-1mm] $-0.0006$\end{tabular}
& $\pm$0.0007 & $\pm$0.001 & $\pm$0.002
\end{tabular}
\end{table*}

\begin{table*}[ht!]
\topcaption{Measured angular coefficients, in bins of \ptmm (in \GeV), for $1 < \abs{\ymm} < 2.4$.}
\label{TableAi2}
\centering
\begin{tabular}{c cccccccc}
\ptmm & 0--10 & 10--20 & 20--35 & 35--55 & 55--80 & 80--120 & 120--200 & 200--400 \\
\hline
$A_0$ & 0.023 & 0.083 & 0.194 & 0.375 & 0.560 & 0.717 & 0.841 & 0.903 \\
Stat. & $\pm$0.002 & $\pm$0.002 & $\pm$0.003 & $\pm$0.003 & $\pm$0.004 & $\pm$0.005 & $\pm$0.008 & $\pm$0.016 \\
Syst. &$\pm$0.010 & $\pm$0.008 & $\pm$0.009 & $\pm$0.007 & $\pm$0.004 & $\pm$0.003 & $\pm$0.004 & $\pm$0.006 \\
\hline
$A_1$ & $-0.0031$ & 0.021 & 0.041 & 0.069 & 0.076 & 0.075 & 0.061 & 0.059 \\
Stat. & $\pm$0.0011 & $\pm$0.001 & $\pm$0.002 & $\pm$0.003 & $\pm$0.004 & $\pm$0.006 & $\pm$0.007 & $\pm$0.013 \\
Syst. & $\pm$0.0008 & $\pm$0.002 & $\pm$0.003 & $\pm$0.004 & $\pm$0.005 & $\pm$0.005 & $\pm$0.004 & $\pm$0.005 \\
\hline
$A_2$ & $-0.002$ & 0.050 & 0.130 & 0.274 & 0.436 & 0.58 & 0.71 & 0.74 \\
Stat. & $\pm$0.002 & $\pm$0.002 & $\pm$0.002 & $\pm$0.003 & $\pm$0.006 & $\pm$0.01 & $\pm$0.02 & $\pm$0.03 \\
Syst. & $\pm$0.010 
& \begin{tabular}{@{}c@{}}$+0.009$ \\[-1mm] $-0.008$\end{tabular}
& \begin{tabular}{@{}c@{}}$+0.004$ \\[-1mm] $-0.005$\end{tabular}
& $\pm$0.007 & $\pm$0.010 & $\pm$0.01 & $\pm$0.01 & $\pm$0.01 \\
\hline
$A_3$ & 0.0017 & 0.0021 & 0.0063 & 0.0108 & 0.022 & 0.035 & 0.062 & 0.070 \\
Stat. & $\pm$0.0008 & $\pm$0.0009 & $\pm$0.0011 & $\pm$0.0016 & $\pm$0.003 & $\pm$0.004 & $\pm$0.007 & $\pm$0.013 \\
Syst. & $\pm$0.0010 & $\pm$0.0009 & $\pm$0.0008 
& \begin{tabular}{@{}c@{}}$+0.0008$ \\[-1mm] $-0.0009$\end{tabular}
& $\pm$0.001 & $\pm$0.002 & $\pm$0.002 & $\pm$0.004 \\
\hline
$A_4$ & 0.059 & 0.054 & 0.052 & 0.047 & 0.038 & 0.034 & 0.025 & 0.029 \\
Stat. & $\pm$0.001 & $\pm$0.002 & $\pm$0.002 & $\pm$0.003 & $\pm$0.003 & $\pm$0.004 & $\pm$0.005 & $\pm$0.010 \\
Syst. & $\pm$0.003 & $\pm$0.002 & $\pm$0.002 
& \begin{tabular}{@{}c@{}}$+0.001$ \\[-1mm] $-0.002$\end{tabular}
& $\pm$0.001 
& \begin{tabular}{@{}c@{}}$+0.002$ \\[-1mm] $-0.001$\end{tabular}
& $\pm$0.002 & $\pm$0.004 \\
\hline
$A_5$ & $<$0.0001 & 0.0011 & 0.0002 & 0.0024 & 0.0034 & 0.003 & 0.008 & 0.015 \\
Stat. & $\pm$0.0009 & $\pm$0.0009 & $\pm$0.0011 & $\pm$0.0016 & $\pm$0.0024 & $\pm$0.003 & $\pm$0.005 & $\pm$0.011 \\
Syst. & $\pm$0.0002 & $\pm$0.0003 & $\pm$0.0003 & $\pm$0.0004 & $\pm$0.0005 & $\pm$0.001 & $\pm$0.002 & $\pm$0.003 \\
\hline
$A_6$ & $-0.0004$ & 0.0016 & 0.0015 & 0.0008 & 0.0077 & 0.0030 & 0.004 & $-0.007$ \\
Stat. & $\pm$0.0010 & $\pm$0.0012 & $\pm$0.0015 & $\pm$0.0020 & $\pm$0.0027 & $\pm$0.0034 & $\pm$0.005 & $\pm$0.010 \\
Syst. & $\pm$0.0003 & $\pm$0.0004 & $\pm$0.0004 & $\pm$0.0005 
& \begin{tabular}{@{}c@{}}$+0.0007$ \\[-1mm] $-0.0008$\end{tabular}
& $\pm$0.0009 & $\pm$0.002 & $\pm$0.003 \\
\hline
$A_7$ & 0.0012 & 0.0017 & 0.0022 & 0.0023 & 0.0050 & 0.0031 & 0.006 & $-0.0005$ \\
Stat. & $\pm$0.0007 & $\pm$0.0008 & $\pm$0.0010 & $\pm$0.0014 & $\pm$0.0020 & $\pm$0.0027 & $\pm$0.004 & $\pm$0.0090 \\
Syst. & $\pm$0.0002 & $\pm$0.0002 & $\pm$0.0002 & $\pm$0.0003 
& \begin{tabular}{@{}c@{}}$+0.0005$ \\[-1mm] $-0.0006$\end{tabular}
& \begin{tabular}{@{}c@{}}$+0.0007$ \\[-1mm] $-0.0008$\end{tabular}
& $\pm$0.001 
& \begin{tabular}{@{}c@{}}$+0.0031$ \\[-1mm] $-0.0034$\end{tabular}
\end{tabular}
\end{table*}

\begin{table*}[ht]
\centering
\topcaption{Measured $A_0-A_2$ difference in bins of \ptmm (in \GeV) and $\abs{\ymm}$.}
\label{TableLT}
\begin{tabular}{c cccccccc}
\ptmm & 0--10 & 10--20 & 20--35 & 35--55 & 55--80 & 80--120 & 120--200 & 200--400 \\
\hline
\multicolumn{9}{c} {$\abs{\ymm}<1$} \\		
$A_{0}-A_{2}$ & 0.035 & 0.039 & 0.077 & 0.130 & 0.150 & 0.18 & 0.15 & 0.17 \\
Stat. & $\pm$0.002 & $\pm$0.003 & $\pm$0.003 & $\pm$0.004 & $\pm$0.006 & $\pm$0.01 & $\pm$0.01 & $\pm$0.03 \\
Syst. & $\pm$0.012 
& \begin{tabular}{@{}c@{}}$+0.012$ \\[-1mm] $-0.011$\end{tabular}
& $\pm$0.011 & $\pm$0.010 & $\pm$0.020 & $\pm$0.02 & $\pm$0.01 & $\pm$0.01 \\
\hline
\multicolumn{9}{c} {$1<\abs{\ymm}<2.4$} \\
$A_{0}-A_{2}$ & 0.024 & 0.034 & 0.064 & 0.100 & 0.120 & 0.14 & 0.13 & 0.16 \\
Stat. & $\pm$0.002 & $\pm$0.003 & $\pm$0.004 & $\pm$0.005 & $\pm$0.006 & $\pm$0.01 & $\pm$0.01 & $\pm$0.03 \\
Syst. & $\pm$0.014 & $\pm$0.012 & $\pm$0.010 & $\pm$0.010 & $\pm$0.010 & $\pm$0.01 & $\pm$0.01 & $\pm$0.01
\end{tabular}
\end{table*}

\cleardoublepage \section{The CMS Collaboration \label{app:collab}}\begin{sloppypar}\hyphenpenalty=5000\widowpenalty=500\clubpenalty=5000\cmsinstitute{Yerevan Physics Institute, Yerevan, Armenia}
{\tolerance=6000
A.~Gevorgyan\cmsorcid{0000-0003-2751-9489}, A.~Hayrapetyan, V.~Makarenko\cmsorcid{0000-0002-8406-8605}, A.~Tumasyan\cmsAuthorMark{1}\cmsorcid{0009-0000-0684-6742}
\par}
\cmsinstitute{Institut f\"{u}r Hochenergiephysik, Vienna, Austria}
{\tolerance=6000
W.~Adam\cmsorcid{0000-0001-9099-4341}, L.~Benato\cmsorcid{0000-0001-5135-7489}, T.~Bergauer\cmsorcid{0000-0002-5786-0293}, M.~Dragicevic\cmsorcid{0000-0003-1967-6783}, P.S.~Hussain\cmsorcid{0000-0002-4825-5278}, M.~Jeitler\cmsAuthorMark{2}\cmsorcid{0000-0002-5141-9560}, N.~Krammer\cmsorcid{0000-0002-0548-0985}, A.~Li\cmsorcid{0000-0002-4547-116X}, D.~Liko\cmsorcid{0000-0002-3380-473X}, M.~Matthewman, J.~Schieck\cmsAuthorMark{2}\cmsorcid{0000-0002-1058-8093}, R.~Sch\"{o}fbeck\cmsAuthorMark{2}\cmsorcid{0000-0002-2332-8784}, M.~Shooshtari\cmsorcid{0009-0004-8882-4887}, M.~Sonawane\cmsorcid{0000-0003-0510-7010}, N.~Van~Den~Bossche\cmsorcid{0000-0003-2973-4991}, W.~Waltenberger\cmsorcid{0000-0002-6215-7228}, C.E.~Wulz\cmsAuthorMark{2}\cmsorcid{0000-0001-9226-5812}
% ...
\par}
$^{1}$Also at Yerevan State University, Yerevan, Armenia\\
$^{2}$Also at Technische Universit\"{a}t Wien, Vienna, Austria\\
% ...
\end{sloppypar}
\end{document}